\RequirePackage{ifpdf}
\ifpdf 
\documentclass[pdftex]{sigma}
\else
\documentclass{sigma}
\fi

\usepackage{mathrsfs}
\usepackage{dsfont}

\newcommand{\Az}{\mathcal{A}_0}
\newcommand{\A}{\mathcal{A}_{\theta}}

\newcommand{\Dt}{\Delta_{\theta}}

\newcommand{\Pa}{\mathbb{C}\mathscr{P}}
\newcommand{\M}{\mathcal{M}}
\newcommand{\dx}{{\rm d}}
\newcommand{\e}{{\rm e}}

\newcommand{\I}{\mathds{1}}

\numberwithin{equation}{section}

\begin{document}

\allowdisplaybreaks

\renewcommand{\thefootnote}{$\star$}

\renewcommand{\PaperNumber}{052}

\FirstPageHeading

\ShortArticleName{Quantum Fields on Noncommutative Spacetimes: Theory and Phenomenology}

\ArticleName{Quantum Fields on Noncommutative Spacetimes:\\ Theory and Phenomenology\footnote{This paper is a
contribution to the Special Issue ``Noncommutative Spaces and Fields''. The
full collection is available at
\href{http://www.emis.de/journals/SIGMA/noncommutative.html}{http://www.emis.de/journals/SIGMA/noncommutative.html}}}

\Author{Aiyalam P. BALACHANDRAN~$^\dag$, Alberto IBORT~$^\ddag$, Giuseppe MARMO~$^\S$\\ and Mario MARTONE~$^{\dag\S}$}

\AuthorNameForHeading{A.P. Balachandran, A.~Ibort, G.~Marmo and M.~Martone}

\Address{$^\dag$~Department of Physics, Syracuse University, Syracuse, NY 13244-1130, USA}
\EmailD{\href{mailto:bal@physics.syr.edu}{bal@physics.syr.edu}, \href{mailto:mcmarton@syr.edu}{mcmarton@syr.edu}}

\Address{$^\ddag$~Departamento de Matem\'aticas, Universidad Carlos III de Madrid,\\
\hphantom{$^\ddag$}~28911 Legan\'es, Madrid, Spain}
\EmailD{\href{mailto:albertoi@math.uc3m.es}{albertoi@math.uc3m.es}}

\Address{$^\S$~Dipartimento di Scienze Fisiche, University of Napoli and INFN,\\
\hphantom{$^\S$}~Via Cinthia I-80126 Napoli, Italy}
\EmailD{\href{mailto:marmo@na.infn.it}{marmo@na.infn.it}}

\ArticleDates{Received March 24, 2010, in f\/inal form June 08, 2010;  Published online June 21, 2010}

\Abstract{In the present work we review the twisted f\/ield construction of quantum f\/ield theory on noncommutative spacetimes based on twisted Poincar\'e invariance. We present the latest development in the f\/ield, in particular the notion of equivalence of such quantum f\/ield theories on a noncommutative spacetime, in this regard we work out explicitly the inequivalence between twisted quantum f\/ield theories on Moyal and Wick--Voros planes; the duality between deformations of the multiplication map on the algebra of functions on spacetime $\mathscr{F}(\mathbb{R}^4)$ and coproduct deformations of the Poincar\'e--Hopf algebra $H\mathscr{P}$ acting on~$\mathscr{F}(\mathbb{R}^4)$; the appearance of a nonassociative product on $\mathscr{F}(\mathbb{R}^4)$ when gauge f\/ields are also included in the picture.
The last part of the manuscript is dedicated to the phenomenology of noncommutative quantum f\/ield theories in the particular approach adopted in this review. CPT violating processes, modif\/ication of two-point temperature correlation function in CMB spectrum analysis and Pauli-forbidden transition in ${\rm Be}^4$ are all ef\/fects which show up in such a noncommutative setting. We review how they appear and in particular the constraint we can infer from comparison between theoretical computations and experimental bounds on such ef\/fects. The best bound we can get, coming from Borexino experiment, is $\gtrsim10^{24}$~TeV for the energy scale of noncommutativity, which corresponds to a length scale $\lesssim 10^{-43}$~m. This bound comes from a dif\/ferent model of spacetime deformation more adapted to applications in atomic physics. It is thus model dependent even though similar bounds are expected for the Moyal spacetime as well as argued elsewhere.}


\Classification{81R50; 81R60}

\renewcommand{\thefootnote}{\arabic{footnote}}
\setcounter{footnote}{0}

\section{Introduction}\label{sec:intro}
A great deal of ef\/fort has been put in trying to achieve a formulation of a theory of quantum gravity.

It has long been speculated that spacetime structure changes radically at Planck scales. In 1994, in a fundamental paper, Doplicher, Fredenhaghen and Roberts argued that the commutator among the four spacetime coordinates of a physical theory in which Einstein's theory of relativity and principles of Quantum Mechanics both coexist, should not vanish\footnote{The idea of extra-dimensions has been widely used recently. Even in those cases the description of spacetime in terms of a four dimensional manifold must be achieved in some sort of ef\/fective limit. It has been shown that the four dimensional limit of certain string theories~\cite{String} does, in fact, involve a noncommutative spacetime.}~\cite{Doplicher}. This argument suggested that noncommutativity of spacetime at Planck scales, a feature seen before as just a way to regularize quantum f\/ield theory, is possibly one of the main features of quantum gravity. We will now sketch the argument, for further details we refer to the original papers~\cite{Doplicher,DFR2}.

From quantum physics we know that to probe a spacetime region with radius of the order of the  Planck length
\begin{gather*}
L_P=\sqrt{\frac{G\hbar}{c^3}}\simeq1.6\times10^{-33}~{\rm cm} ,
\end{gather*}
we need a particle of mass $M^*$ such that its Compton wavelength is smaller than the length scale of the spacetime region we wish to probe, namely of the Planck length:
\begin{gather*}
\lambda_C=\frac{\hbar}{M^*c}\leq L_P\quad\Rightarrow\quad M^*\geq\frac{\hbar}{L_Pc}\simeq10^{19}~{\rm Gev}.
\end{gather*}

Einstein's theory of relativity tells us that the Schwarzschild radius associated to such a mass distribution is
\begin{gather*}
R_S=\frac{2GM^*}{c^2}\geq 2L_P .
\end{gather*}
It is thus greater than the region we would like to explore. Thus probing spacetime at the Planck scale generates a paradox: in the process an event horizon (or a trapped surface) is created which now prevents us to access altogether the spacetime region we were initially interested in.

In order to avoid the collapse of the probed region, we must assume that it is not possible to simultaneously measure all four spacetime coordinates. This requirement can be incorporated in the noncommutativity of coordinates. A simple choice for this noncommutativity is
\begin{gather} \label{UV1}
[\widehat{x}_{\mu}, \widehat{x}_{\nu}] = i \theta_{\mu \nu}.
\end{gather}
Here $\theta_{\mu \nu} = - \theta_{\nu \mu}$ are constants and
$\widehat{x}_{\mu}$ are the coordinate functions on an $n$-dimensional spacetime $\mathbb{R}^n$ on which the Poincar\'e group acts in a standard manner for $\theta_{\mu\nu}=0$:
\[
\widehat{x}_{\mu}(x) = x_{\mu}.
\]

In the present paper, the relations (\ref{UV1}) will be assumed. We will describe how to formulate the minimal requirements for a quantum f\/ield theory (QFT) on such a spacetime and two particular instances of quantum f\/ield theories will be constructed over such noncommutative spacetimes, the Moyal and the Wick--Voros quantum f\/ields.  We should point it out here that dif\/ferent proposals have been discussed in the literature (see for instance \cite{Lizzi,Vitale} and refe\-ren\-ces therein), however the construction provided here is constructive and their consistency and properties can be tested directly.  The paper will be organised as follows. In the next section we will introduce the Drinfel'd twist and the deformation of the Poincar\'e Hopf algebra $H\mathscr{P}$. Both concepts will play a crucial role in the formulation of QFT on noncommutative spacetime, which will be then explained in Section~\ref{sectionIII}. Relations (\ref{UV1}) do not uniquely specify the spacetime algebra. For example, both Moyal and Wick--Voros planes are compatible with~(\ref{UV1}). In Sections~\ref{sectionIV} and~\ref{sectionV} we will see how this freedom gets ref\/lected on the QFT side. In Section~\ref{sectionVI} we will present some further mathematical developments while the last section will be devoted to describe phenomenological consequences of noncommutative spacetimes. They inf\/luence the cosmic microwave background, the $K^0$--$\bar{K}^0$ mass dif\/ference and the Pauli principle among \mbox{others}. From available data we will present the experimental bounds on the scale of spacetime noncommutativity. The best bound we f\/ind for its energy scale is $\gtrsim10^{24}$~{\rm TeV}. It comes from Borexino and Superkamiokande experiments. We will then conclude with some f\/inal remarks.

\section{The Drinfel'd twist and deformed coproduct}\label{sectionII}

At f\/irst we notice that the relations (\ref{UV1}) can be implemented by deforming the product of the standard commutative algebra of functions on our $n$-dimensional space-time $\Az\equiv(\mathscr{F}(\mathcal{M}),m_0$), where ${\cal M}\cong \mathbb{R}^{n}$, $\mathscr{F}(\M)$ are smooth functions on $\M$ and $m_0$ is the point-wise multiplication map:
\begin{gather*}
m_{0} (f \otimes g)(x)=f(x)g(x)=g(x)f(x)=m_{0} (g\otimes f)(x).
\end{gather*}
There is a general procedure to deform such a product in a controlled way using the so-called {\it twist deformation} \cite{drinfeld}.

Let us denote as $\mathcal{A}_\theta=(\mathscr{F}(\M),m_\theta)$ such a deformation of $\mathcal{A}_0$ which leads to (\ref{UV1}). Here $m_\theta=m_0\circ\mathcal{F}_\theta$ is a noncommutative product and $\mathcal{F}_\theta$ the {\it twist map}. The two satisfy:
\begin{gather}\label{sig1}
 m_\theta(f\otimes g)(x)=m_0\circ\mathcal{F}_\theta(f\otimes g)(x):=(f\star g)(x) ,  \\
 \mathcal{F}_\theta: \ \mathscr{F}(\M)\otimes\mathscr{F}(\M)\to\mathscr{F}(\M)\otimes\mathscr{F}(\M)\qquad{\rm and}\qquad\mathcal{F}_\theta\to\I\quad {\rm as}\quad \theta\to 0 .\nonumber
\end{gather}
An explicit form of $\mathcal{F}_\theta$ is
\begin{gather}\label{twi1}
\mathcal{F}_{\theta}=\exp\frac{i}{2}\theta[\partial_x\otimes \partial_y-\partial_y\otimes \partial_x].
\end{gather}
In particular the unit is preserved by the deformation. We notice that
\begin{itemize}\itemsep=0pt
\item[1)] $\mathcal{F}_\theta$ is one-to-one and invertible;

\item[2)] $\mathcal{F}_\theta$ acts on the tensor product in a non-factorizable manner, i.e.\ the action on $\mathscr{F}(\M)\otimes\mathscr{F}(\M)$ intertwines the two factors.
\end{itemize}

The above choice of $\mathcal{F}_\theta\equiv\mathcal{F}^\M_\theta$ leads to the Moyal plane $\mathcal{A}^\M_\theta$:
\begin{gather*}
m^\M_\theta(f\otimes g)(x,y)\equiv f(x,y)\cdot g(x,y)+\frac{i}{2}\theta\big[(\partial_xf)(\partial_yg)-(\partial_yf)(\partial_xg)\big]\\
\phantom{m^\M_\theta(f\otimes g)(x,y)\equiv}{}
+\sum_{n=2}\frac{\big[\frac{i}{2}\theta(\partial_x\otimes\partial_y-\partial_y\otimes\partial_x)\big]^n}{n!}(f\otimes g).
\end{gather*}
But it is not unique. Another choice, leading to the Voros plane $\A^V$ is
\begin{gather}\label{twi2}
\mathcal{F}^{V}_{\theta}=\exp\frac{1}{2}\theta[\partial_x\otimes \partial_x+\partial_y\otimes \partial_y]\mathcal{F}^{\mathcal{M}}_{\theta}=\mathcal{F}^\M_\theta\exp\frac{1}{2}\theta[\partial_x\otimes \partial_x+\partial_y\otimes \partial_y]  .
\end{gather}

The noncommutative relations  (\ref{UV1}) bring with them another problem: at f\/irst sight it seems that the noncommutativity of spacetime coordinates violates Poincar\'e invariance: the l.h.s.\ of~(\ref{UV1}) transforms in a non-trivial way under the standard action of the Poincar\'e group $\mathscr{P}$ whereas the r.h.s.\ does not. But there is still a way to act properly with the Poincar\'e group, or rather, with its group algebra $\mathbb{C}\mathscr{P}$, on the deformed algebras $\A$ if its action is changed to the so-called twisted action \cite{chaichian,wess,sasha}. It goes as follows. The l.h.s.\ of~(\ref{UV1}) is an element of the tensor product space followed by $m_\theta:\A^{\M,V}\otimes\A^{\M,V}\to\A^{\M,V}$. The way in which $\mathbb{C}\mathscr{P}$ acts on the tensor product space requires a homomorphism $\mathbb{C}\mathscr{P}\to\mathbb{C}\mathscr{P}\otimes\mathbb{C}\mathscr{P}$. It is not f\/ixed by the way $\mathscr{P}$ acts on $\A$, but is a further ingredient of the theory we need to specify. This map is called the coproduct and will be denoted by $\Delta$:
\[
\Delta : \ \mathbb{C}\mathscr{P}\to\mathbb{C}\mathscr{P}\otimes\mathbb{C}\mathscr{P}.
\]
Once we provide $\mathbb{C}\mathscr{P}$ with the coproduct $\Delta$, that is we also specify how $\mathbb{C}\mathscr{P}$ acts on a tensor product space, we get a new mathematical structure called a Hopf algebra $H\mathscr{P}$. There are further formal compatibility requirements between the multiplication map of $\mathbb{C}\mathscr{P}$ and the map~$\Delta$, but we will not discuss them here. For further details on Hopf algebras, we refer to~\cite{Dito,chari,majid,aschieri}.

Usually we assume for the coproduct the simple separable map $\Delta_0$:
\begin{gather}\label{sig2}
\Delta_0(g)=g\otimes g\qquad\forall \, g\in\mathscr{P} .
\end{gather}
It extends to $\mathbb{C}\mathscr{P}$ by linearity. But (\ref{sig2}) is not the only possible choice. The idea proposed in~\cite{chaichian,wess,sasha} is that we can assume a dif\/ferent coproduct on $\mathbb{C}\mathscr{P}$, that is ``twisted'' or deformed with respect to $\Delta_0$, to modify the action of the Poincar\'e group on tensor product spaces in such a way that it does preserve relations (\ref{UV1}). We must realize here that the change of $\Delta_0$ is not a mere mathematical construction, as it af\/fects the way composite systems transform under spacetime symmetries.  This observation will have deep consequences in the physical interpretation of the theory as it will be shown later. This modif\/ication changes the standard Hopf algebra structure associated with the Poincar\'e group (the Poincar\'e--Hopf algebra $H\mathscr{P}$) to a twisted Poincar\'e--Hopf algebra $H_{\theta}\mathscr{P}$ ($H_0\mathscr{P}\equiv H\mathscr{P}$).

The deformed algebra is not unique. For the Moyal and Wick--Voros cases they are dif\/ferent, although isomorphic\footnote{The two deformations are in fact equivalent in Hopf algebra deformation theory. That is they belong to the same equivalence class in the non-Abelian cohomology that classif\/ies Hopf algebra twist-deformations. See for details~\cite{majid}.}. The isomorphism map is
\begin{gather*}
\gamma=\e^{-\frac{1}{4}\theta(\partial_x^2+\partial_y^2)},\qquad\gamma: \ \mathcal{A}_\theta^\M\to \mathcal{A}_\theta^V\qquad{\rm and}\qquad\mathcal{F}^V_\theta=\gamma\otimes\gamma\mathcal{F}^\M_\theta\Delta\big(\gamma^{-1}\big) .
\end{gather*}
We denote them by $H_{\theta}^{\M,V}\mathscr{P}$ when we want to emphasise that we are working with Moyal and Wick--Voros spacetimes.

Still they lead to dif\/ferent QFT's since the isomorphism map is not unitary. We discuss this point in detail in Sections~\ref{sectionIV} and~\ref{sectionV}.

The explicit form of the deformed coproduct $\Delta_\theta$ of $H_{\theta}\mathscr{P}$ is obtained from requiring that the action of $\mathbb{C}\mathscr{P}$ is an {\it automorphism} of the new algebra of functions $\A$ on spacetime. That is, the action of $\mathbb{C}\mathscr{P}$ has to be compatible with the new noncommutative multiplication rule~(\ref{sig1}). In particular, for $g\in\mathscr{P}\subset\mathbb{C}\mathscr{P}$,
\begin{gather*}
g\triangleright m_\theta(f\otimes h)(x)=m_\theta(g\triangleright (f\otimes h))(x)    .
\end{gather*}

It is easy to see that the standard coproduct choice (\ref{sig2}) is not compatible \cite{chaichian,wess,sasha} with the action of $\mathbb{C}\mathscr{P}$ on the deformed algebra $\A$. In the cases under consideration, where $\A$ are twist deformations of $\mathcal{A}_0$, there is a simple rule to get deformations $\Dt$ of $\Delta_0$ compatible with $m_{\theta}$. They are given by the formula:
\begin{gather}\label{UV3}
\Dt=(F_{\theta})^{-1}\Delta_0 F_{\theta},
\end{gather}
where $F_{\theta}$ is an element in $H_{\theta}\mathscr{P}\otimes H_{\theta}\mathscr{P}$ and it is determined by the map $\mathcal{F}_\theta$ introduced before, $\mathcal{F}_\theta$ being the realisation of $F_\theta$ on $\A\otimes\A$.

For $\mathcal{F}_\theta=\mathcal{F}_\theta^{\M,V}$, the corresponding $F_\theta^{\M,V}$ give us the Hopf algebras $H^{\M,V}_\theta\mathscr{P}$.

Without going deeper into the deformation theory of Hopf algebras, we just note that the deformations we are considering here are very specif\/ic ones since we keep the multiplication rules unchanged and deform only the co-structures of the underlying Hopf algebra. Thus for $\mathbb{C}\mathscr{P}$, we only change $\Delta_0$ to $\Dt$ leaving the group multiplication the same. For a deeper discussion on deformations of algebras and Hopf algebras, we refer again to the literature \cite{Dito,chari,majid,aschieri}.

Lastly we have to introduce the concept of twisted statistics. It is a strict consequence of the twisted action of the Poincar\'e group on the tensor product space (\ref{UV3}). In quantum mechanics two kinds of particles, with dif\/ferent statistics, are admitted: fermions, which are described by fully antisymmetrised states, and bosons, which are instead completely symmetric. Let $\mathcal{H}$ be a~single particle Hilbert space. Then given a two-particle quantum state, $\alpha\otimes\beta$ with $\alpha,\beta\in\mathcal{H}$, we can get its symmetrised and anti-symmetrised parts as:
\[
\alpha\otimes_{S,A}\beta=\frac{\I\pm\tau_0}{2}\alpha\otimes\beta,
\]
where the map $\tau_0$ is called the {\it flip operator} and it simply switches the elements on which it acts,
\[
\tau_0(\alpha\otimes\beta)=\beta\otimes\alpha.
\]

From the foundations of quantum f\/ield theories it can be proved that the statistics of particles have to be superselected, that is Poincar\'e transformations cannot take bosons (fermions) into fermions (bosons). In other words, a symmetric (antisymmetric) state must still be symmetric (antisymmetric) after the action of any element of the Poincar\'e group. This requirement implies that the f\/lip operator has to commute with the coproduct of any element of $\mathbb{C}\mathscr{P}$. As can be trivially checked, the action of $\tau_0$ commutes with the coproduct $\Delta_0(g)$ of $g\in\mathscr{P}$, but not with~$\Delta_\theta(g)$. If we do not modify the f\/lip operator, we end up with a theory in which, for example, a~rotation can transform a fermion into a boson.

If the deformation of the coproduct is of the kind we have been considering so far, that is a~twist deformation as in (\ref{UV3}), again it is easy to f\/ind a deformation $\tau_\theta$ of the f\/lip operator $\tau_0$ which commutes with $\Delta_\theta$:
\begin{gather}\label{flip}
\tau_0   \to  \tau_\theta=(F_\theta)^{-1}\tau_0F_\theta
\end{gather}
and, moreover $\tau_\theta^2 = I$.
This equation contains the $\mathcal{R}$-matrix of a quasi-triangular Hopf algebra. By def\/inition, $\mathcal{R}$ is given by
\begin{gather}\label{sig5}
\tau_\theta=\mathcal{R}\circ\tau_0 .
\end{gather}
Hence:
\begin{gather}\label{sig3}
\mathcal{R}:=F^{-1}_\theta\circ F_{\theta 21},\qquad F_{\theta21}\equiv\tau_0F_\theta\tau_0^{-1} .
\end{gather}
In quantum physics on a noncommutative spacetime, we then consider symmetrisation (antisymmetrisation) with respect to $\tau_\theta$ rather then $\tau_0$:
\begin{gather}\label{sig6}
\alpha\otimes_{S_\theta,A_\theta}\beta=\frac{\I\pm\tau_\theta}{2}\alpha\otimes\beta .
\end{gather}

\section[Examples: Moyal and Wick-Voros planes]{Examples: Moyal and Wick--Voros planes}\label{sectionIII}

The preceding section explains the general features of the formalism of deformations of algebras induced by the relations~(\ref{UV1}) and how to deform the Hopf algebra structure of the Poincar\'e group to keep the theory still invariant under its action, as in~(\ref{UV3}). We now proceed with the study of  two explicit examples: Moyal and Wick--Voros planes.

As previously indicated, we denote by $\mathcal{F}_{\theta}^{\mathcal{M},V}$ and $m_{\theta}^{\mathcal{M},V}$ the twists and multiplication maps for the deformed algebras $\A^{\mathcal{M},V}$ respectively.  Then as in (\ref{sig1})
\begin{gather}\label{UV2}
m_{\theta}^{\mathcal{M},V}(f\otimes g)\equiv m_0\circ\mathcal{F}_{\theta}^{\mathcal{M},V}(f\otimes g) .
\end{gather}

In the following, for the sake of simplicity, we will work in two dimensions. The generalization to arbitrary dimensions will be discussed in Section~\ref{sectionVII}.

In two dimensions, we can  always write $\theta_{\mu\nu}$ as
\begin{gather}\label{UV10}
\theta_{\mu\nu} = \theta\epsilon_{\mu\nu},\qquad
\epsilon_{01}= - \epsilon_{10}=1,
\end{gather}
where $\theta$ is a constant. Then as in (\ref{twi1}) and (\ref{twi2}) the two twists $\mathcal{F}_{\theta}^{\mathcal{M},V}$ assume the form
\begin{gather}
\mathcal{F}^{\mathcal{M}}_{\theta}=\exp\frac{i}{2}\theta[\partial_x\otimes \partial_y-\partial_y\otimes \partial_x] ,\nonumber\\
\label{UV9}
 \mathcal{F}^{V}_{\theta}=\exp\frac{1}{2}\theta[\partial_x\otimes \partial_x+\partial_y\otimes \partial_y]\mathcal{F}^{\mathcal{M}}_{\theta}=\mathcal{F}^\M_\theta\exp\frac{1}{2}\theta[\partial_x\otimes \partial_x+\partial_y\otimes \partial_y]  .
\end{gather}

Hereafter we will call $\cal{F}^{\cal{M}}_{\theta}$ and $\mathcal{F}_{\theta}^{V}$ the Moyal and Wick--Voros twists, and $\A^{\mathcal{M}}$ and $\A^{V}$ the Moyal and Wick--Voros algebras respectively. Both deformations, $\A^{\mathcal{M}}$ and $\A^V$, realize the commutation relations (\ref{UV1}). This fact already shows, as pointed out above, how noncommutativity of spacetime does not f\/ix uniquely the deformation of the algebra. There are many more noncommutative algebras of functions on spacetime that realize~(\ref{UV1}) with dif\/ferent twisted products. In order to address the study of how this freedom ref\/lects on the quantum f\/ield theory side it is enough to work with two of them. Thus hereafter we will only work with $\A^{\mathcal{M},V}$.

Given the above expressions for the twists, explicit expressions for the noncommutative pro\-duct of the functions in the two cases follow immediately from (\ref{UV2}):
\begin{gather}\label{Moya}
(f \star_{\mathcal{M}} g)(x)=m^\M_{\theta} (f\otimes g)(x)\equiv f(x)\textrm{e}^{\frac{i}{2}\theta_{\alpha \beta}\overleftarrow{\partial^{\alpha}}\otimes \overrightarrow{\partial^{\beta}}} g(x),\\
 (f \star_V g)(x)=m^V_{\theta} (f\otimes g)(x) \equiv f(x)\textrm{e}^{\frac{i}{2}\big(\theta_{\alpha \beta}\overleftarrow{\partial^{\alpha}}\otimes \overrightarrow{\partial^{\beta}} -i\theta\delta_{\alpha\beta}\overleftarrow{\partial^{\alpha}}\otimes \overrightarrow{\partial^{\beta}}\big)}g(x).
 \nonumber
\end{gather}
If we let the $\star$-product to act on the coordinate functions, we get in both cases the noncommutative relations (\ref{UV1}).

Once the two twists are given, following (\ref{UV3}), we can immediately write down the deformations of the two coproducts as well:
\begin{gather}\label{sig7}
\Delta^{\M,V}_\theta(g)=\big(F^{\M,V}_\theta\big)^{-1}\Delta_0(g)F^{\M,V}_\theta=\big(F^{\M,V}_\theta\big)^{-1}(g\otimes g)F^{\M,V}_\theta.
\end{gather}
Here $F^{\M,V}_\theta\in H\mathscr{P}\otimes H\mathscr{P}$ are:
\begin{gather}\label{Moya1}
F^{\mathcal{M}}_{\theta}=\exp\left(-\frac{i}{2}\theta[P_x\otimes P_y-P_y\otimes P_x]\right) ,\\
F^{V}_{\theta}=\exp\left(-\frac{1}{2}\theta[P_x\otimes P_x+P_y\otimes P_y]\right)F^{\mathcal{M}}_{\theta}=F^\M_\theta\exp\left(-\frac{1}{2}\theta[P_x\otimes P_x+P_y\otimes P_y]\right)   ,\nonumber
\end{gather}
where $P_\mu$ are translation generators. Their realisation on $\mathcal{A}^{\M,V}_\theta$ is:
\begin{gather*}
P_\mu\triangleright f(x)\equiv(P_\mu f)(x)=-i(\partial_\mu f)(x) .
\end{gather*}

We end this section with a discussion on how the $\tau_0$ gets twisted in the two cases. As $F^{\M}_\theta$ is skew-symmetric and $F^V_\theta$ is the composition of $F^\M_\theta$ and a symmetric part, we get:
\begin{gather*}
 F^\M_{\theta21}=\big(F^\M_\theta\big)^{-1} ,\\
 F^V_{\theta21}=\exp\left(-\frac{1}{2}\theta[P_x\otimes P_x+P_y\otimes P_y]\right)\big(F^\M_\theta\big)^{-1} .
\end{gather*}

In the $\mathcal{R}$-matrix (\ref{sig3}), any symmetric part of the twist cancels. Thus in both Moyal and Wick--Voros cases, the statistics of particles is twisted in the same way:
\[
\mathcal{R}^{\M,V}=\big(F^\M_\theta\big)^{-2}=\exp\left(-i\theta[P_x\otimes P_y-P_y\otimes P_x]\right) .
\]

We must point out here that the results discussed in this paper dif\/fer from those in~\cite{Lizzi} because of the dif\/ferences in the construction of quantum f\/ield theories on $\A^{\M,V}$ here and in~\cite{Vitale}. Thus while our approach is based on Hilbert spaces, operators and explicitly enforces unitarity, the other approach uses functional integrals. There are issues to be clarif\/ied in the case of functional integrals in the context of the Moyal plane as they do not fulf\/ill ref\/lection positivity. For a detailed discussion of this issue, see~\cite{bondia}.

\section{Quantum f\/ield theories on Moyal and Voros planes}\label{sectionIV}

It is time now to discuss how to quantize the two theories introduced in the previous section. Our approach to the Moyal plane is discussed in \cite{sasha,mangano} and to the Wick--Voros plane can be found in \cite{Mario,Mario3}. For another approach to the latter, see \cite{Lizzi}.

The quantization procedure consists in f\/inding a set of creation operators, and by adjointness the annihilation counterpart, which create multiparticle states providing a unitary representation of the twisted Poincar\'e--Hopf symmetry. Such a set of creation and annihilation operators must also implement the appropriate twisted statistics (\ref{sig5})--(\ref{sig6}). Out of them we can construct the twisted quantum f\/ields (QFs). It has been in fact proven elsewhere \cite{sasha,mangano} that the Hamiltonian constructed out of such twisted f\/ields is Hopf--Poincar\'e invariant.

Let us f\/irst consider the Moyal case. As our previous work \cite{sasha,mangano} shows,
\begin{gather}\label{new1}
 a^\M_p=c_p\exp\left(-\frac{i}{2}p_\mu\theta^{\mu\nu}P_\nu\right),\\
\label{new2}
 a^{\M\dag}_p=c^\dag_p\exp\left(\frac{i}{2}p_\mu\theta^{\mu\nu}P_\nu\right),
\end{gather}
where $c_p$, $c^\dag_p$ are the untwisted $\theta^{\mu\nu}=0$ annihilation and creation operators (we can assume all such operators to refer to in-, out- or free-operators as the occasion demands), provide the operators we were looking for. Let's see that.

Since we are considering only deformations in which the coproduct is changed, the way in which the Poincar\'e group acts on a single particle state is the usual one. Therefore the creation operator (\ref{new2}) on the vacuum will act like the untwisted operator $c^\dag_p$. It is then plausible that the generators of Poincar\'e transformations on the Hilbert space under consideration have to be the untwisted ones. We will now conf\/irm this: if $(a,\Lambda)\to U(a,\Lambda)\equiv U(\Lambda)U(a)$ is the $\theta=0$ unitary representation of the Poincar\'e group, we will show that the multiparticle states created by acting with (\ref{new2}) on the vacuum transform with the Moyal coproduct (\ref{sig7}).

If we consider the action of a general group element of the Poincar\'e group $(a,\Lambda)$ on a two-particle state $|p,q\rangle_\theta$, we expect
\begin{gather}
\Delta^\M_\theta\big((a,\Lambda)\big)\triangleright|p,q\rangle_\theta = \big(F^\M_\theta\big)^{-1}\big((a,\Lambda)\otimes (a,\Lambda)\big)F^\M_\theta\triangleright|p,q\rangle_\theta\nonumber\\
\phantom{\Delta^\M_\theta\big((a,\Lambda)\big)\triangleright|p,q\rangle_\theta}{}
 = |\Lambda p,\Lambda q\rangle_\theta\e^{-\frac{i}{2}(p\wedge q-\Lambda p\wedge\Lambda q)}\e^{-i(p+q)\cdot a},\label{sig14}
\end{gather}
where $a\wedge b:=a_\mu\theta^{\mu\nu}b_\nu$, and we have used the properties of a momentum eigenstate, $P_\mu\triangleright|p\rangle=p_\mu|p\rangle$ and $(a,\Lambda)\triangleright|p\rangle=|\Lambda p\rangle\e^{-ip\cdot a}$.

For the $\theta=0$ generators of the unitary representation of the Poincar\'e group on the Hilbert space under consideration, we have
\[
U(a,\Lambda)c^\dag_pU^{-1}(a,\Lambda)=\e^{ip\cdot a}c^\dag_{\Lambda p}.
\]
Def\/ining \cite{BaPiQue}\footnote{In the case under study, because of the twisted statistics, the creation operators, and likewise their adjoints, do not commute. The order in which they act on a state becomes then an issue. The choice made here is motivated by asking for consistency~\cite{sasha}. The scalar product we consider for the def\/inition of the adjoint is the one associated with the untwisted creation and annihilation operators.}
\begin{gather}\label{sig25}
|p,q\rangle^\M_\theta=a^{\M\dag}_qa^{\M\dag}_p|0\rangle
\end{gather}
we can now explicitly compute how $U(a,\Lambda)$ acts on the two-particle state considered in (\ref{sig25}):
\begin{gather}
U(a,\Lambda)|p,q\rangle_\theta = U(a,\Lambda)c^\dag_qU^{-1}(a,\Lambda)U(a,\Lambda)\e^{\frac{i}{2}q\wedge P}c^\dag_p|0\rangle=c^\dag_{\Lambda q}c^\dag_{\Lambda p}|0\rangle\e^{-i(p+q)\cdot a}\e^{\frac{i}{2}q\wedge p}\nonumber\\
\phantom{U(a,\Lambda)|p,q\rangle_\theta }{}
= a^{\M\dag}_{\Lambda q}\e^{\frac{i}{2}q\wedge P}c^\dag_{\Lambda p}|0\rangle\e^{\frac{i}{2}q\wedge p}\e^{-i(p+q)\cdot a} = |\Lambda p,\Lambda q\rangle\e^{-\frac{i}{2}(p\wedge q-\Lambda p\wedge \Lambda q)}\e^{-i(p+q)\cdot a}.\label{sig15}
\end{gather}
In the above computation we have used $P_\nu|0\rangle=0$ and a relation which will be used repeatedly in what follows: $\e^{\frac{i}{2}p_\mu\theta^{\mu\nu}P_\nu}c^\dag_q\e^{-\frac{i}{2}p_\mu\theta^{\mu\nu}P_\nu}=\e^{\frac{i}{2}p_\mu\theta^{\mu\nu}[P_\nu,\, \cdot\, ]}c^\dag_q=\e^{\frac{i}{2}p_\mu\theta^{\mu\nu}q_\nu}c^\dag_q$. Thus~(\ref{sig15}) coincides with~(\ref{sig14}). It is a remarkable fact that the appropriate deformation of the coproduct naturally appears as the $\theta=0$ Poincar\'e group generator acts on the two particle states obtained by the creation operator~(\ref{new2}). This result generalises to $n$-particles states.

The two-particle state in (\ref{sig14}) also fulf\/ills the twisted statistics. From (\ref{sig5}), (\ref{sig6}) we expect that (throughout this paper we will only consider the bosonic case):
\[
|p,q\rangle^\M_\theta=\frac{\I+\tau^\M_\theta}{2}|p,q\rangle=\frac{1}{2}\big(|p,q\rangle+\e^{-i\vec{q}\wedge \vec{p}}|q,p\rangle\big)
\]
and
\[
|q,p\rangle^\M_\theta=\frac{\I+\tau^\M_\theta}{2}|q,p\rangle=\frac{1}{2}\big(|q,p\rangle+\e^{-i\vec{p}\wedge \vec{q}}|p,q\rangle\big)=\e^{-i\vec{q}\wedge \vec{p}}|p,q\rangle^\M_\theta.
\]
On the other hand using the def\/inition (\ref{sig25}) of two-particle states in terms of  the creation operators $a^{\dag\M}_p$'s:
\begin{gather*}
|q,p\rangle^\M_\theta=a^{\M\dag}_pa^{\M\dag}_q|0\rangle=c^\dag_p\exp\left(\frac{i}{2}p_\mu\theta^{\mu\nu}P_\nu\right)c^\dag_q|0\rangle\\
\phantom{|q,p\rangle^\M_\theta}{}
=\e^{-\frac{i}{2}\vec{q}\wedge\vec{p}}a^\dag_q\exp(-\frac{i}{2}q_\mu\theta^{\mu\nu}P_\nu)c^\dag_p|0\rangle
=\e^{-i\vec{q}\wedge\vec{p}}|p,q\rangle_\theta,
\end{gather*}
where we have used the relation introduced above and the fact that $c^\dag_p$ and $c^\dag_q$ commute. So these creation operators do implement the statistics we want.

The operators def\/ined in (\ref{new1}), (\ref{new2}) are called in the literature \cite{Grosse, Zamo, Faddeev} as {\it dressed ope\-ra\-tors}. They are obtained from the $\theta=0$ ones by {\it dressing} them using the exponential term. Exploiting a peculiar property of the Moyal plane, which will be explained in the next section, we can obtain the Hermitian quantum f\/ield $\varphi_\theta^{\mathcal{M}}$ on the Moyal plane:
\begin{gather}\label{sig10}
\varphi^\M_\theta(x)=\int\dx\mu(p)\big[a^\M_p\e_{-p}(x)+a^{\M\dag}_p\e_{p}(x)\big],
\end{gather}
where $\e_p(x)$ denotes $\e^{ip\cdot x}$ as usual, through a similar dressing of the standard $\theta=0$ scalar f\/ield:
\begin{gather}\label{new3}
\varphi^\M_\theta=\varphi_0\e^{\frac{1}{2}\overleftarrow{\partial}_\mu\theta^{\mu\nu}P_\nu} .
\end{gather}
This formula is f\/irst deduced for in-, out- or free-f\/ields. For example,
\[
\varphi^{\M,{\rm in}}_\theta=\varphi_0^{{\rm in}}\e^{\frac{1}{2}\overleftarrow{\partial}_\mu\theta^{\mu\nu}P_\nu} .
\]
But since the Heisenberg f\/ield becomes the `in' f\/ield as $x_0\to-\infty$,
\[
\varphi_0(x)\to\varphi^{{\rm in}}_0\qquad{\rm as}\quad x_0\to-\infty ,
\]
and $P_\mu$ is time-independent, we (at least heuristically) infer (\ref{new3}) for the fully interacting Heisenberg f\/ield.

Products of the f\/ield (\ref{sig10}) have a further remarkable property which we have called {\it self-reproducing property}:
\[
\big(\varphi_\theta^\M\star_\M\varphi^\M_\theta\big)(x)=\big[(\varphi_0\cdot\varphi_0)(x)\big]
\e^{\frac{1}{2}\overleftarrow{\partial}_\mu\theta^{\mu\nu}P_\nu} ,
\]
where the $\cdot$ represents the standard point-wise product. This property generalises to products of $N$ f\/ields
\begin{gather}\label{sig12}
\underbrace{\varphi^\M_\theta\star_\M\varphi^\M_\theta\star_\M \cdots \star_\M\varphi^\M_\theta}_{N\text{-factors}}=\varphi_0^N\e^{\frac{1}{2}\overleftarrow{\partial}_\mu\theta^{\mu\nu}P_\nu},
\end{gather}
where again $\varphi_0^N$ indicates the $N$-th power with respect the commutative product $m_0$. This self-reproducing property plays a signif\/icant role in general theory. It is the basis for the proof of the absence of UV-IR mixing in Moyal f\/ield theories (with no gauge f\/ields) \cite{BaPiQur,BaPiQue}.

Now consider the Wick--Voros case. The twisted creation operators, which correctly create states from the vacuum transforming under the twisted coproduct, are \cite{Mario}
\begin{gather}\label{new4}
a^{V\dag}_p=c^\dag_p\e^{\frac{i}{2}(p_\mu\theta^{\mu\nu}P_\nu-i\theta p_\nu P_\nu)},
\end{gather}
where $p_\nu P_\nu$ uses the Euclidean scalar product. Its adjoint is
\begin{gather}\label{new5}
a^V_p=\e^{-\frac{i}{2}(p_\mu\theta^{\mu\nu}P_\nu+i\theta p_\nu P_\nu)}c_p .
\end{gather}
We prove elsewhere \cite{BaIbMaMa} that (\ref{new4}) and (\ref{new5}) are also dictated by the covariance of quantum f\/ields.

It can be shown that, like in Moyal case, the states obtained by the action of $a^{V\dag}_\theta$ reproduce the appropriate twisted statistics too.

Although we have obtained, like in Moyal, (\ref{new4}) and (\ref{new5}) by dressing the $\theta=0$ operators, the quantum f\/ield $\varphi_\theta^{V}$ on Wick--Voros plane:
\begin{gather}\label{sig11}
\varphi^V_\theta(x)=\int\dx\mu(p)\big[a^V_p\e_{-p}(x)+a^{V\dag}_p\e_{p}(x)\big]
\end{gather}
cannot be obtained from an overall dressing like in (\ref{new3}). This property fails due to the fact that is not possible to factorise the same overall exponential out of both (\ref{new4}) and (\ref{new5}), since it is not possible in (\ref{new4}), (\ref{new5}) to move the exponential from right (left) to left (right), that is:
\begin{gather}
a^{V\dag}_p\neq\exp\left(-\frac{i}{2}(p_\mu\theta^{\mu\nu}P_\nu+i\theta p_\nu P_\nu)\right)c^\dag_p ,\nonumber\\
\label{sig16}
a^V_p\neq c_p\exp\left(-\frac{i}{2}(p_\mu\theta^{\mu\nu}P_\nu+i\theta p_\nu P_\nu)\right) .
\end{gather}
This is because $c^\dag_p$ and $c_p$ do not commute with the exponentials in (\ref{new4}) and (\ref{new5}), in fact moving $c^\dag_p$ ($c_p$) to the right (left) will bring a factor $\e^{-\frac{\theta}{2}p_\nu p_\nu}$:
\begin{gather*}
a^{V\dag}_p=\exp\left(\frac{i}{2}(p_\mu\theta^{\mu\nu}P_\nu-i\theta p_\nu P_\nu)-\frac{\theta}{2}p_\nu p_\nu\right)c^\dag_p,\\
a^{V}_p=c_p\exp\left(-\frac{i}{2}(p_\mu\theta^{\mu\nu}P_\nu+i\theta p_\nu P_\nu)-\frac{\theta}{2}p_\nu p_\nu\right).
\end{gather*}
A consequence is that we have to twist the creation-annihilation parts $\varphi^{(\pm)I}_0$ ($I={}$ in-, out- or free-) f\/ields separately:
\begin{gather*}
\varphi_\theta^{(+)V,{\rm I}}=\int\dx\mu(p)a^{V,{\rm I}\dag}\e_p=\varphi_0^{(+){\rm I}}e^{\frac{1}{2}(\overleftarrow{\partial}_\mu\theta^{\mu\nu}P_\nu-i\theta\overleftarrow{\partial}_\mu P_\mu)},\\
\varphi_\theta^{(-)V,{\rm I}}=\int\dx\mu(p)a^{V,{\rm I}}\e_{-p}=e^{\frac{1}{2}(\overrightarrow{\partial}_\mu\theta^{\mu\nu}P_\nu+i\theta\overrightarrow{\partial}_\mu P_\mu)}\varphi_0^{(-){\rm I}},\\
\e_p(x)=\e^{ip\cdot x},\qquad\dx\mu(p):=\frac{\dx^3p}{2\sqrt{\vec{p}\,{}^2+m^2}},\qquad m={\rm mass\ of\ the\ f\/ield}\ \varphi_0^{\rm I},
\end{gather*}
where now we have added the superscript I to $\varphi^{\rm I}_0$, $a^{V,{\rm I}\dag}_p$, and $a^{V,{\rm I}}_p$.

Therefore to obtain the f\/ield (\ref{sig11}) we have to twist creation and annihilation parts separately
\begin{gather}\label{sig13}
\varphi^{V,{\rm I}}_\theta=\varphi^{(+)V,{\rm I}}_\theta+\varphi^{(-)V,{\rm I}}_\theta.
\end{gather}

A further point relates to the self-reproducting property of these Wick--Voros f\/ields. The quantum f\/ields $\varphi^{(\pm){\rm I}V}_\theta$ are self-reproductive, but in dif\/ferent ways. Thus
\begin{gather*}
\underbrace{\varphi^{(+)V,{\rm I}}_\theta\star_V\varphi^{(+)V,{\rm I}}_\theta\star_V \cdots \star_V\varphi^{(+)V,{\rm I}}_\theta}_{M\text{-factors}}=\big(\varphi^{(+){\rm I}}_0\big)^M\e^{\frac{1}{2}(\overleftarrow{\partial}_\mu)\theta^{\mu\nu}P_\nu-i\theta\overleftarrow{\partial}_\mu P_\mu},\\
\underbrace{\varphi^{(-)V,{\rm I}}_\theta\star_V\varphi^{(-)V,{\rm I}}_\theta\star_V\cdots \star_V\varphi^{(-)V,{\rm I}}_\theta}_{M'\text{-factors}}=\e^{\frac{1}{2}(\overrightarrow{\partial}_\mu)\theta^{\mu\nu}P_\nu+i\theta\overrightarrow{\partial}_\mu P_\mu}\big(\varphi^{(-){\rm I}}_0\big)^{M'},
\end{gather*}
where, as in (\ref{sig12}), $\big(\varphi^{(\pm){\rm I}}_0\big)^N$ is the $N$-th power of $\varphi_0^{(\pm)I}$ with respect to the commutative pro\-duct~$m_0$. Given (\ref{sig13}), it follows that the full f\/ield, $\varphi^V_\theta$, does not have the self-reproducing property.

\section{Voros versus Moyal: a comparison}\label{sectionV}

It is now time to compare the two theories we have previously introduced. We will follow the treatment in \cite{Mario3} and prove them to be inequivalent. First we want to approach this question of equivalence from a heuristic point of view. We will conclude this section proving that no similarity transformation can relate the two theories, showing their inequivalence with respect to isospectral transformations.

The f\/irst big dif\/ference among QFT's on Moyal and Wick--Voros planes has already appeared above. Although we have not stressed it, (\ref{new1}) and (\ref{new2}) could have been changed to
\begin{gather} \label{Ago2}
a^{\mathcal{M}}_p=\exp\left({-\frac{i}{2}p_{\mu}\theta^{\mu\nu}P_{\nu}}\right)c_p,
\\
\label{Ago3}
a^{\mathcal{M}\dagger}_p=\exp\left({\frac{i}{2}p_{\mu}\theta^{\mu\nu}P_{\nu}}\right)c_p^{\dagger}.
\end{gather}
But in fact (\ref{new1}) equals (\ref{Ago2}) and (\ref{new2}) equals (\ref{Ago3}) because $\theta^{\mu\nu}=-\theta^{\nu\mu}$ \cite{mangano,vaidya}. In (\ref{sig16}) we have instead seen that this is not the case for the Wick--Voros plane. This observation is important since it ensures that the Moyal f\/ield (\ref{sig10}), which we can obtain from the $\theta=0$ f\/ield by twisting it, is Hermitian. Had we constructed the Voros f\/ield in a similar way, we would have ended up with a non-Hermitian operator.

Equation (\ref{sig16}) is also the reason for the absence of the self-reproducing property on the Wick--Voros plane that by itself causes dif\/ferences at the level of physics. (For example the arguments for the absence of UV/IR mixing on the Moyal plane would fail here.)

But that is not all. The states in the Wick--Voros case are not normalised in the same way as in the Moyal case. For instance
\begin{gather}
 \langle k_1,k_2|p_1,p_2\rangle\equiv\langle0|a^{V,{\rm I}}_{k_1}a^{V,{\rm I}}_{k_2}a^{V,{\rm I}\dag}_{p_2}a^{V,{\rm I}\dag}_{p_1}|0\rangle =\e^{\theta k_1\cdot k_2}4\sqrt{(\vec{k}_1^2+m^2)(\vec{k}_2^2+m^2)}\nonumber
  \\\label{long3}
  \hphantom{\langle k_1,k_2|p_1,p_2\rangle\equiv}{}\times
 \big[\delta^3(k_1-p_1)\delta^3(k_2-p_2)+\e^{\frac{i}{2}k_{1\mu}\theta^{\mu\nu} k_{2\nu}}\delta^3(k_1-p_2)\delta^3(k_2-p_1)\big],\\
 {\rm I= in,\ out,}\qquad|0\rangle_{\rm `in}=|0\rangle_{\rm out},\qquad m={\rm mass\ of\ the\ f\/ield}\ \varphi_0^I .
 \nonumber
\end{gather}

For scattering theory, normalisation is important. It depends on the scalar product we are using. If we normalise the states as in the Moyal case, since the normalisation constant in~(\ref{long3}) is momentum dependent, the normalised states no longer transform with the Wick--Voros coproduct.  The factor~(\ref{long3}) has been computed using the standard scalar product in the Fock space. We can try changing it~\cite{Mario} so that the states become correctly normalised. But then the representation $(a,\Lambda)\to U(a,\Lambda)$ ceases to be unitary. We can try to seek for another one, modifying $U(a,\Lambda)$ using a non trivial operator $\mathcal{O}$. We could not f\/ind any such $\mathcal{O}$ which would preserve the way the unitary representation has to act on single particle states. It then appears that the Wick--Voros plane is not suitable for quantum physics.

We now show that there is no similarity transformation taking $a^{\M,{\rm I}}_p$, $a^{\M,{\rm I}\dag}_p$ into $a^{V,{\rm I}}_p$, $a^{V,{\rm I}\dag}_p$. One way to quickly see this is to examine the operators without the Moyal part of the twist. So we consider $c^{{\rm I}}_p$, $c^{{\rm I}\dag}_p$ and
\begin{gather}\label{new7}
a^{V,{\rm I}'}_p=\e^{\frac{1}{2}\theta p_\nu P_\nu}c^{\rm I}_p ,\\
a^{V,{\rm I}'\dag}_p=c^{{\rm I}\dag}_p\e^{\frac{1}{2}\theta p_\nu P_\nu} .\label{new8}
\end{gather}

Now
\begin{gather*}
[c^{\rm I}_p,c^{{\rm I}\dag}_k]=2|p_0|\delta^3(p-k)\I,\\
p_0=\sqrt{\vec{p}\,{}^2+m^2},\qquad m={\rm mass\ of\ the\ f\/ield}\ \varphi_0^{\rm I} .
\end{gather*}

Had there existed an invertible operator $W$ such that
\[
Wc^{\rm I}_pW^{-1}=a^{V,{\rm I}'}_p,\qquad Wc^{{\rm I}\dag}_pW^{-1}=a^{V,{\rm I}'\dag}_p ,
\]
then we would have
\begin{gather}\label{sig30}
[a^{V,{\rm I}'}_p,a^{V,{\rm I}'\dag}_k]=2|p_0|\delta^3(p-k)\I .
\end{gather}
But a direct calculation of the l.h.s.\ using (\ref{new7}), (\ref{new8}) shows that (\ref{sig30}) is not true.

However there exists an invertible operator $S$ which transforms $a^{\M,{\rm I}\dag}_p$ to $a^{V,{\rm I}\dag}_p$ and is given by:
\begin{gather} \label{new9}
 S=\e^{\frac{\theta}{4}(P_\mu P_\mu+ K)}, \qquad K=-\int\dx\mu(k)k_\mu k_\mu c^{{\rm I}\dag}_kc^{\rm I}_k
\end{gather}
and a simple computation now shows that
\[
Sa^{\M,{\rm I}\dag}_pS^{-1}=a^{V,{\rm I}\dag}_p.
\]
where, as usual, I on $c^{{\rm I}\dag}_k$, $c^{\rm I}_k$ denotes in-, out- or free- while in $P_\mu P_\mu$ and $k_\mu k_\mu$, we use the Euclidean scalar product.

But
\[
Sa^{\M,{\rm I}}_pS^{-1}=\e^{-\frac{i}{2}(p_\mu\theta^{\mu\nu}P_\nu-i\theta p_\nu p_\nu)}c^{\rm I}_p=\tilde{a}^{V,{\rm I}}_p\neq a^{V,{\rm I}}_p.
\]

Let us pursue the properties of this operator further.

The operator $S$ leaves the vacuum invariant and shows that certain correlators in the Moyal and Wick--Voros cases are equal. From the explicit expression (\ref{new9}) it follows also that the map induced by the operator $S$ is isospectral, but not unitary in the Fock space scalar product. It is possible to def\/ine a new scalar product which makes $S$ unitary~\cite{Mario}. But the previously def\/ined $U(a,\Lambda)$ would not be unitary in this scalar product as we discussed above.

Now consider the twisted f\/ields
\begin{gather*}
 \varphi^{\M,{\rm I}}_\theta=\int\dx\mu(p)\big[a^{\M,{\rm I}}_p\e_{-p}+a^{\M,{\rm I}\dag}_p\e_{p}\big] , \\
 \tilde{\varphi}^{V,{\rm I}}_\theta=\int\dx\mu(p)\big[\tilde{a}^{V,{\rm I}}_p\e_{-p}+a^{V,{\rm I}\dag}_p\e_{p}\big] ,
\end{gather*}
where $\e_p(x)$ denotes $\e^{ip\cdot x}$ as before. Then of course,
\begin{gather}\label{new10}
S: \ \varphi^{\M,{\rm I}}_\theta\to S\triangleright\varphi^{\M,{\rm I}}_\theta:=\int\dx\mu(p)S\big[a^{\M,{\rm I}}_p\e_{-p}+a^{\M,{\rm I}\dag}_p\e_p\big]S^{-1}=\tilde{\varphi}^{V,{\rm I}}_\theta .
\end{gather}
Also
\begin{gather}\label{new11}
S|0\rangle=S^{-1}|0\rangle=0.
\end{gather}

From (\ref{new10}) and (\ref{new11}) we obtain trivially the equality of the $n$-points correlation functions:
\[
\langle\varphi^{\M,{\rm I}}_\theta(x_1)\varphi^{\M,{\rm I}}_\theta(x_2)\cdots \varphi^{\M,{\rm I}}_\theta(x_N)\rangle_0=\langle\tilde{\varphi}^{V,{\rm I}}_\theta(x_1)\tilde{\varphi}^{V,{\rm I}}_\theta(x_2)\cdots \tilde{\varphi}^{V,{\rm I}}_\theta(x_N)\rangle_0 .
\]

Consider simple interaction densities such as
\begin{gather*}
\mathscr{H}^\M_{\rm I}=\underbrace{\varphi^{\M,{\rm I}}_\theta\star_\M\varphi^{\M,{\rm I}}_\theta\star_\M\cdots \star_\M\varphi^{\M,{\rm I}}_\theta}_{N\text{-factors}}\qquad{\rm and}\qquad\mathscr{H}^V_{\rm I}=\underbrace{\tilde{\varphi}^{V,{\rm I}}_\theta\star_V\tilde{\varphi}^{V,{\rm I}}_\theta\star_V\cdots \star_V\tilde{\varphi}^{V,{\rm I}}_\theta}_{N\text{-factors}}
\end{gather*}
in either f\/ield. Since $S$ only acts on the operator parts of the f\/ields, the similarity transformation in (\ref{new10}) will not map $\mathscr{H}^\M_{\rm I}$ to $\mathscr{H}^V_{\rm I}$:
\[
S\triangleright\mathscr{H}^\M_{\rm I}\neq\mathscr{H}_{\rm I}^V .
\]
Hence
\begin{gather*}
 \langle\varphi^{\M,{\rm I}}_\theta(x_1)\cdots \varphi^{\M,{\rm I}}_\theta(x_j)\mathscr{H}^\M_{\rm I}(x_{j+1})\varphi^{\M,{\rm I}}_\theta(x_{j+2})\cdots \varphi^{\M,{\rm I}}_\theta(x_N)\rangle_0\\
\qquad{} \neq\langle\tilde{\varphi}^{V,{\rm I}}_\theta(x_1)\cdots \tilde{\varphi}^{V,{\rm I}}_\theta(x_j)\mathscr{H}^V_{\rm I}(x_{j+1})\cdots \tilde{\varphi}^{V,{\rm I}}_\theta(x_{j+2})\cdots \tilde{\varphi}^{V,{\rm I}}_\theta(x_N)\rangle_0 .
\end{gather*}
So we can immediately conclude that also in this case the two theories are dif\/ferent.

There is no such $S$ for mapping $\varphi^{\M,{\rm I}}_\theta$ to $\varphi^{V,{\rm I}}_\theta$, so that the correlators are not equal even at the tree level.

\section{Further formal developments}\label{sectionVI}

In the presentation we have given so far, the Hopf algebra deformation has been introduced to compensate for the spacetime noncommutativity (\ref{UV1}) and restore the Poincar\'e invariance of the theory. The algebra of functions on spacetime and the Poincar\'e group are related by a duality in a very precise manner. The consequence of that is that once one of the two is deformed, the duality relation imposes a similar deformation on the other. A deformation of the coproduct of the Poincar\'e group thus induces a noncommutative spacetime \cite{Mario2}. Thus while customarily, we deduce the Hopf algebra deformations from spacetime noncommutativity, we can also deduce the latter from the former.

Also, so far we have not discussed gauge theory to any extent. In the approach to gauge theories followed in \cite{gauge1,gauge2}, a unique feature seems to arise due to the deformation needed in the presence of gauge f\/ields: the coproduct deformation is not coassociative. Because of the dual relation cited above, the product between functions on spacetime induced by such a~noncoassociative coproduct is also nonassociative \cite{Nonass}.

Both of these more formal features deserve to be discussed in more detail.

\subsection{Spacetime from symmetry}

The starting point to unveil the duality that ties the Hopf algebra deformation with the deformation on the spacetime algebra, is to recall a well-known result. The commutative algebra of functions on spacetime, what we have called $\mathcal{A}_0\equiv(\mathscr{F}(\mathbb{R}^4),m_0)$, can be obtained as a coset of the algebra of functions on the Poincar\'e group with respect to the right action of the proper ortochronous Lorentz group $\mathcal{L}^\uparrow_+$. This result exploits the fact that the Poincar\'e group is the semi-direct product of translations and the Lorentz group, so that $\mathbb{R}^4\cong\mathscr{P}\diagup\mathcal{L}^\uparrow_+$. It is crucial to stress that the reason why such a coset description of $\mathbb{R}^4$ gives the algebraic structure of $\mathcal{A}_0$ is because the product of two right-invariant functions under the action of $\mathcal{L}^\uparrow_+$ on the Poincar\'e group it is still right invariant.

In order to generalise such a construction to the noncommutative case, we should also mention that it is well known in Hopf algebra theory that the product structure on functions on any Lie group $\mathscr{F}(\mathscr{G})$ and the group algebra $\mathbb{C}\mathscr{G}$ are dual to each other. Such a duality is a remarkably powerful tool. Given the Hopf algebra structure on the group algebra $\mathbb{C}\mathscr{G}$, that is given the multiplication map $m$ and the coproduct $\Delta$, and the duality map:
\begin{gather}\label{dua3}
\forall \, f\in\mathscr{F}(\mathscr{G}),\quad \forall \, g\in\mathbb{C}\mathscr{G}\qquad\langle f,g\rangle=f(g)\in\mathbb{C}
\end{gather}
we can deduce a multiplication map $m^*$ and a coproduct $\Delta^*$ on $\mathscr{F}(\mathscr{G})$ turning $\mathscr{F}(\mathscr{G})$ also into a Hopf algebra. The structures on $\mathscr{F}(\mathscr{G})$, $(m^*,\Delta^*)$, are uniquely deduced in the following way:
\begin{gather}
\label{dua1}
\forall \, f_1,f_2\in\mathscr{F}(\mathscr{G}),\ \forall \, g\in\mathbb{C}\mathscr{G},\qquad (f_1\cdot f_2)(g)\equiv\langle m^*(f_1\otimes f_2),g\rangle:=(f_1\otimes f_2)(\Delta(g)) , \\
 \forall \, f\in\mathscr{F}(\mathscr{G}),\ \forall \, g_1,g_2\in\mathbb{C}\mathscr{G}  ,\qquad \Delta(f)(g_1\otimes g_2)\equiv\langle \Delta^*(f),g_1\otimes g_2\rangle:=f(g_1\cdot g_2) .\nonumber
\end{gather}

In this approach, a co-commutative $\Delta$ on $\mathbb{C}\mathscr{G}$, like the choice (\ref{sig2}), induces a commutative multiplication map on functions. It is then natural to expect that if we deform the costructure of $\mathbb{C}\mathscr{P}$, like in (\ref{UV3}), then on the dual, that is on the functions on Poincar\'e group, we will obtain a noncommutative multiplication map.

We will work with $H^\M_\theta\mathscr{P}$. Its coproduct is obtained from the standard coproduct on $\mathbb{C}\mathscr{P}$ using the Moyal twist introduced in (\ref{Moya1}). In order to get an algebra of functions on spacetime from $\mathscr{F}(\mathscr{P})$, as explained above, we should make sure that the coset operation is compatible with the noncommutative product induced by (\ref{dua1}) when $\Delta\equiv\Delta_\theta^\M$. Such a requirement will f\/ix how the duality pairing (\ref{dua3}) gets lifted to the tensor product space. Once this lift is f\/ixed, there is no more freedom allowed and we can proceed to compute the multiplication map $m^*_\theta$ on the algebra $\mathscr{F}(\mathscr{P})$ using (\ref{dua1}). In \cite{Mario2} we have found such a lift and we have explicitly shown that such a product $m^*_\theta$ coincides exactly with the standard Moyal product on $\mathscr{F}(\mathbb{R}^4)$ $m^\M_\theta$ as in (\ref{Moya}).

\subsection{Nonassociative deformations}

Gauge theories in noncommutative QFT's have been a long standing problem. The main complication is the non-closure of the commutator of covariant derivatives if connections are noncommutative. Thus:
\begin{gather*}
[{\bf A_\mu},{\bf A_\nu}](x)=[A_\mu^a\lambda^a,A_\nu^b\lambda^b](x)\\
\phantom{[{\bf A_\mu},{\bf A_\nu}](x)}{}
=\big(A_\mu^a\star A^b_\nu-A_\nu^b\star A^a_\mu\big)(x) \lambda^a\lambda^b+\big(A_\mu^a\star A_\nu^b\big)(x)f^{ab}_c\lambda^c,\qquad \lambda^i\in\mathfrak{g},
\end{gather*}
where $\mathfrak{g}$ is the Lie algebra. Now $\lambda^a\lambda^b\notin\mathfrak{g}$, unless the Lie group $\mathscr{G}$ is $U(N)$ where $N$ is the dimension of the particular representation under consideration. The Lie group in this way becomes the whole unitary group if $\lambda^a$'s describe an irreducible representation. Its dimension is  also dependent on the representation of $\lambda^a$. One way out is to assume standard point-wise multiplication among gauge f\/ields so that the extra term, proportional to $[A_\mu^a,A_\nu^b]_\star(x)$, vanishes as it usually does. We will follow this approach. More formally we will consider connections as $\mathscr{G}$-valued functions on the commutative algebra $\mathcal{A}_0\equiv(\mathscr{F}(\mathbb{R}^4),m_0)$, where $\mathscr{G}$ is the gauge group. The f\/ields $A^\theta_\mu$ are then not twisted: $A^\theta_\mu=A^0_\mu$. Such a formulation, where matter f\/ields are twisted, that is regarded as elements of $\A\equiv(\mathscr{F}(\mathbb{R}^4),m_\theta)$, while gauge f\/ields are treated as standard commutative functions, can be shown to be internally consistent. The basic reason is that $\A$ is an $\mathcal{A}_0$ module.  We refer to~\cite{gauge1,gauge2} for more details.

In this approach, the deformation of the action of the Poincar\'e--Hopf algebra on tensor product spaces shows up only through matter f\/ields. In the particular case in which we choose the Moyal plane $\A\equiv\A^\M$ for the matter part, the deformed coproduct $\overline{\Delta}_\theta$ which acts both on~$\varphi^\M_\theta$ and~$A^\mu_\theta=A^\mu_0$ has to fulf\/ill
\begin{gather*}
 \overline{\Delta}_\theta|_{\rm Gauge\ f\/ields}=\Delta_0 , \\
 \overline{\Delta}_\theta|_{\rm Matter\ f\/ields}=\Delta^\M_\theta=\big(F_\theta^{\M}\big)^{-1}\Delta_0F^\M_\theta .
\end{gather*}

In order to write an expression for $\overline{\Delta}_\theta$ which would take care of both gauge and matter sections, we need to enlarge the group algebra $\Pa$. Specif\/ically we should introduce a central element $u$ which will act ef\/fectively as a grading operator for the quantum f\/ields. We call the extended group algebra $\overline{\Pa}$.

Let us denote by $\chi_0^{g,m}$ a generic untwisted gauge and matter f\/ield. The central element $u$ acts on them by conjugation, we will denote such an action as usual as ${\rm Ad}$:
\[
{\rm Ad}(u)\chi_0^{g,m}:=u\chi^{g,m}_0u^{-1} .
\]
We def\/ine the action of $u$ so that gauge f\/ields are even whereas matter f\/ields are odd:
\[
{\rm Ad}(u)\chi_0^g=\chi_0^g,\qquad
{\rm Ad}(u)\chi_0^m=-\chi_0^m .
\]
It then follows that the operator
\begin{gather}\label{Bab1}
\delta_{{\rm Ad}\,u,-\I}\equiv\frac{1}{2}[\I-{\rm Ad}(u)]
\end{gather}
acts as identity on matter f\/ields $\chi^m_0$ and as 0 on gauge ones $\chi^g_0$.

We have now introduced enough formalism to present the def\/inition of the extended co\-pro\-duct~$\overline{\Delta}_\theta$:
\begin{gather}\label{Bab2}
\overline{\Delta}_\theta:=\mathfrak{F}_\theta^{-1}\Delta_0\mathfrak{F}_\theta,
\end{gather}
where now the twist operator $\mathfrak{F}_\theta$ is given by
\[
\mathfrak{F}_\theta=\e^{-\frac{i}{2}P_\mu\theta^{\mu\nu}\otimes P_\nu(\delta_{{\rm Ad}\, u,-\I}\otimes\I)} .
\]
Using (\ref{Bab1}), $\mathfrak{F}_\theta$ can be checked to reduce to the identity map on the gauge sector and the standard Moyal twist $F^\M_\theta$ (\ref{Moya1}) on matter f\/ields.

We want to study the coproduct deformation introduced in (\ref{Bab2}) further. Specif\/ically we want to study its so-called coassociativity. A coproduct $\Delta$ is said to be coassociative if the following equation is fulf\/illed:
\begin{gather}\label{Bab3}
(\Delta\otimes\I)\Delta(g)=(\I\otimes\Delta)\Delta(g),\qquad\forall\,  g\in\Pa .
\end{gather}
Both l.h.s.\ and r.h.s.\ def\/ine an action of $\Pa$ on a three particles state. If the coproduct is coassociative there is no ambiguity in the def\/inition of the action of $\Pa$ on three, and actually on any multiparticle state. Coassociativity of $\Delta$ is assumed when we deal with Hopf algebras. The lack of it will change a Hopf algebra to what is called a quasi-Hopf algebra (see for instance~\cite{majid}).

In order to show that (\ref{Bab2}) is not coassociative, we can compute the action both of  l.h.s.\ and r.h.s.\ of~(\ref{Bab3}), with $\Delta\equiv\overline{\Delta}_\theta$, on vectors $\e_p\otimes\e_q\otimes\e_k\in V_{\rm Gauge}\otimes V_{\rm Matter}\otimes V_{\rm Matter}$. $V_{\rm Gauge}$ and $V_{\rm Matter}$ denote the even and odd vector spaces under $u$ and $\e_r$ $(r=p,q,k)$ denote plane wave vectors: $\e_r(x):=\e^{ir\cdot x}$. In \cite{Nonass} it has been shown that $(\overline{\Delta}_\theta\otimes\I)\overline{\Delta}_\theta$ and $(\I\otimes\overline{\Delta}_\theta)\overline{\Delta}_\theta$ transform $\e_p\otimes\e_q\otimes\e_k$ in dif\/ferent ways. Thus $\overline{\Delta}_\theta$ is not coassociative and $\overline{H_\theta\mathscr{P}}$, that is the deformation of the Hopf--Poincar\'e algebra when gauge f\/ields also are included in the picture, is a quasi-Hopf algebra.

Given the deformation $\overline{H_\theta\mathscr{P}}$, the deformation on the algebra of functions $\mathscr{F}(\mathbb{R}^4)$ can be obtained using duality as explained previously. It is a standard result that the multiplication map~$m^*$ induced on the dual Hopf algebra is associative if and only if the starting $\Delta$ is coassociative. In $\overline{H_\theta\mathscr{P}}$, the coproduct we need to consider is $\overline{\Delta}_\theta$. The multiplication map on functions on spacetime $\overline{\star}$ induced by (\ref{dua1}) is hence nonassociative:
\[
f_1,  f_2,  f_3\in\mathscr{F}\big(\mathbb{R}^4\big),\qquad (f_1\overline{\star}f_2)\overline{\star}f_3\neq f_1\overline{\star}(f_2\overline{\star}f_3)  .
\]

This is a striking result which deserves more study. In particular one can try to understand how to formulate quantum mechanics on such nonassociative spacetimes.

\section[Experimental bounds on $\theta$]{Experimental bounds on $\boldsymbol{\theta}$}\label{sectionVII}

In the following we will present some bounds for the noncommutativity parameter $\theta$ provided by the most recent experiments.   Most of them deal with QFT on the Moyal plane.

\subsection{Modif\/ications of the CMB spectrum}

The Cosmic Microwave Background (CMB) radiation gives us a picture of the universe when it was only 400,000 years old. The smallness of the anisotropies detected by COBE in 1992 have suggested that what we see today is how the universe appeared at the end of an era of exponential expansion called inf\/lation. The inf\/lation era stretched a region of Planck size into cosmological scales. We then expect that the CMB radiation would show some signatures of physics at Planck scale too. That is the reason why descriptions of CMB radiation in the noncommutative framework have been addressed in many places \cite{NCMB1,NCMB2,NCMB3,NCMB4,NCMB5,NCMB6,NCMB7,NCMB8,NCMB9,NCMB10}. We will report the results of \cite{CMB1,CMB2} which use the approach to QFT on noncommutative spaces introduced above. We will only report the most important formulas and results, referring to the original papers for details. The $\theta=0$ calculation can be found in \cite{CMBt1,CMBt2,CMBt3}.

The observational data used to constrain the noncommutative parameter $\theta$ were obtained from WMAP5 \cite{WMAP1,WMAP2,WMAP3}, ACBAR \cite{ACb1,ACb2,ACb3} and CBI \cite{CB1,CB2,CB3,CB4,CB5}. All these experiments have measured the anisotropy in the temperature of the CMB.

The temperature f\/luctuations can be expanded in spherical harmonics:
\[
\frac{\Delta T(\hat{n})}{T}=\sum_{lm}a_{lm}Y_{lm}(\hat{n}),
\]
where $\hat{n}$ is the direction of the incoming photons.

The coef\/f\/icients of spherical harmonics, $a_{lm}$, contain all the information encoded in the temperature f\/luctuations. They can be connected to the primordial scalar metric perturbation caused by the quantum f\/luctuations of the inf\/laton f\/ield:
\[
a^{0,\theta}_{lm}=4\pi(-i)^l\int\frac{\dx^3k}{(2\pi)^3}\Delta_l(k)\varphi_{0,\theta}(\vec{k},\eta)Y^*_{lm}(\hat{k}),
\]
where $\Delta_l(k)$ are called {\it transfer functions}, $\eta=\frac{t}{a(t)}$ is the cosmological time and $\varphi_{0,\theta}(\vec{k},\eta)$ is the inf\/laton f\/ield in momentum space which for $\theta\neq0$ is taken to be given by the Moyal quantum f\/ield~(\ref{sig10}). We can now introduce the two-point temperature correlation function, which can also be expanded in spherical harmonics:
\[
\left\langle\frac{\Delta T(\hat{n})}{T}\frac{\Delta T(\hat{n}')}{T}\right\rangle=\sum_{lm,l'm'}\langle a_{lm}a^*_{l'm'}\rangle Y^*_{lm}(\hat{n})Y_{l'm'}(\hat{n}'),
\]
where
\begin{gather}\label{sig17}
\langle a_{lm}a^*_{l'm'}\rangle_{0,\theta}=16\pi^2(-i)^{l-l'}\int\frac{\dx^3k}{(2\pi)^3}\Delta_l(k)
\Delta_{l'}(k)P_{\varphi_{0,\theta}}(\vec{k})Y^*_{lm}(\hat{k})Y_{l'm'}(\hat{k}).
\end{gather}

The function $P_{\varphi_{0,\theta}}(\vec{k})$ is called {\it power spectrum} and is def\/ined in terms of the real part of the two-point correlation function in momentum space:
\[
\langle\varphi_{0,\theta}(\vec{k},\eta)\varphi^\dag_{0,\theta}(\vec{k}',\eta)\rangle
=(2\pi)^3P_{\varphi_{0,\theta}}(\vec{k})\delta^3(\vec{k}-\vec{k}').
\]
It is $P_{\varphi_{0,\theta}}(\vec{k})$ which contains the signature of noncommutativity. The expression for the power spectrum when a mode $k$ crosses the horizon, that is when $a(\eta)H=k$ say for $\eta=\eta_0$, in the commutative limit and in the $\theta\neq0$ cases, are respectively:
\begin{gather} \label{sig18}
P_{\varphi_0}(\vec{k})=\frac{16\pi G}{9 \epsilon}\frac{H^2}{2k^3}\Big|_{a(\eta_0)H=k},\\
\label{sig19}
P_{\varphi_\theta}(\vec{k})=P_{\varphi_0}(\vec{k})\cosh\big(H\vec{\theta}\,{}^0\cdot \vec{k}\big).
\end{gather}
These formulas above are obtained assuming the Hubble parameter $H$ to be constant du\-ring inf\/lation, $\epsilon$ is the slow-roll parameter in the single f\/ield inf\/lation model~\cite{CMBt1} and $\vec{\theta}^0:=(\theta^{01},\theta^{02},\theta^{03})$.

Substituting (\ref{sig18}) and (\ref{sig19}) in (\ref{sig17}) we get for the two cases:
\begin{gather} \label{sig20}
\langle a_{lm}a^*_{l'm'}\rangle_0 = \frac{2}{\pi}\int\dx k\ k^2\ (\Delta_l(k))^2P_{\varphi_0}(k)\delta_{ll'}\delta_{mm'}\equiv C_l\delta_{ll'}\delta_{mm'}\\
\nonumber
\langle a_{lm}a^*_{l'm'}\rangle_\theta = \frac{2}{\pi}\int\dx k\sum^\infty_{l''=0,\, l'':{\rm even}}(i)^{l-l'}(-1)^m(2l''+1)k^2\Delta_l(k)\Delta_{l'}(k)P_{\varphi_0}(k)i_{l''}(\theta kH)\\
\label{sig21}
\phantom{\langle a_{lm}a^*_{l'm'}\rangle_\theta =}{} \times\sqrt{(2l+1)(2l''+1)}
\left(\begin{array}{ccc}
l&l'&l''\\
0&0&0
\end{array}\right)
\left(\begin{array}{ccc}
l&l'&l''\\
-m&m'&0
\end{array}\right),
\end{gather}
where $i_l(z)$ is the modif\/ied spherical Bessel function, while the Wigner's 3-$j$ symbols come from integrals of spherical harmonics:
\begin{gather*}
\int\dx\Omega_kY_{l,-m}(\hat{k})Y_{l',m'}(\hat{k})Y_{l'',0}(\hat{k})
=\sqrt{\frac{(2l{+}1)(2l'{+}1)(2l''{+}1)}{4\pi}}
\left(\!\begin{array}{ccc}
l&l'&l''\\
0&0&0
\end{array}\!\right)\!
\left(\!\begin{array}{ccc}
l&l'&l''\\
-m&m'&0
\end{array}\!\right)\! .
\end{gather*}

From (\ref{sig20}) we get the def\/inition of $C_l$, that is the power spectrum for multipole moment $l$. The data sets from WMAP5, ACBAR and CBI are only available for such diagonal terms. We can then consider (\ref{sig21}) for $l=l'$ and average over $m$ to get a much simpler expression:
\begin{gather}\label{sig22}
C^\theta_l\equiv\frac{1}{2l+1}\sum_m\langle a_{lm}a^*_{lm}\rangle_\theta=\int\dx k\ k^2\ P_{\varphi_\theta}(k)|\Delta_l(k,\eta=\eta_0)|^2i_0(\theta kH) .
\end{gather}
Here the coef\/f\/icients $C^0_l\equiv C_l$ in the $\theta=0$ are given by (\ref{sig20}).

A comparison between (\ref{sig20}) and (\ref{sig22}) with the available data shows \cite{CMB2} that it is not possible to constrain $\theta$ using WMAP data. This is due to the fact that the last data point in the WMAP observations is at a relatively small value for angular momentum $l$ ($l=839$) whereas the signature of noncommutativity appears at very high values of~$l$. Using data from ACBAR and CB instead, which go up to $l=2985$ and $l=3500$ respectively, we can get constraints on~$\theta$. The restriction we found on $H\theta$ is
\begin{gather}\label{sig23}
H\theta<0.01~{\rm Mpc} .
\end{gather}
Data from ACBAR+WMAP3, give an upper limit on the Hubble parameter:
\begin{gather}\label{sig24}
H<1.704  \times  10^{-5}M_p ,
\end{gather}
where $M_p$ is the Planck mass. We are interested in a value for $\theta$ at the end of inf\/lation. The scale factor then, assuming a reheating temperature of the universe close to the GUT energy ($10^{16}$~GeV), is $a\simeq10^{-29}$~\cite{CMBt1}.  Using both (\ref{sig23}) and (\ref{sig24}) we f\/inally get a bound for $\theta$:
\[
\sqrt{\theta}<\big(1.84  a \times  10^{-9}\big)^{1/2}=1.36  \times  10^{-19}~{\rm m},
\]
which corresponds to a lower bound for the energy scale of 10~TeV.

\subsection{Pauli-forbidden transitions}

There have been suggestions \cite{vaidya} that the Pauli principle will be modif\/ied on non-commutative spacetimes. In \cite{pot}, the authors computed the statistical potential $V_{\rm STAT}$ among two twisted two-particle fermion states. $V_{\rm STAT}$ does not show an inf\/initely repulsive core, as it would be on commutative spacetimes. In \cite{Pramod} a concrete calculation of Pauli-forbidden transition has been carried out on a non-commutative spacetime. It leads to the bound $\gtrsim10^{24}$~TeV for the energy scale of noncommutativity when combined with limits on Pauli-forbidden transitions from \cite{expe1,expe2,expe3,expe4,expe5}. This is a strong bound suggesting an energy scale beyond the Planck scale.

The calculation is carried out using a twist deformation which is neither Moyal nor Wick--Voros. Specif\/ically the twist has the form:
\begin{gather}\label{PP1}
\mathcal{F}'_{\theta\hat{n}}:=\e^{-\frac{i\theta}{2}(P_0\otimes\hat{n}\cdot\vec{J}-\hat{n}\cdot\vec{J}\otimes P_0)},
\end{gather}
where $P_0$ and $\vec{J}$ are respectively the time translation and rotations generators and $\hat{n}$ is a f\/ixed unit vector. The twist (\ref{PP1}) leads to a dif\/ferent kind of noncommutativity with respect to (\ref{UV1}). Acting with the noncommutative product $m'_\theta:=m_0\circ\mathcal{F}'_{\theta\hat{n}}$ on coordinates functions we get
\[
[\hat{x}_0,\hat{x}_j]=i\theta\epsilon_{ijk}n_i\hat{x}_k .
\]

It can be shown that the twist (\ref{PP1}) still def\/ines a consistent Poincar\'e--Hopf algebra deformation. It is then possible to proceed to compute the twisted ground two-electron state of a Be atom:
\[
|1,1\rangle_{\theta\hat{n}}=\frac{\I-\tau_{\theta\hat{n}}}{\sqrt{2}}|1,+1;1,-1\rangle_{\hat{n}},
\]
where $\tau_{\theta\hat{n}}=\big(\mathcal{F}'_{\theta\hat{n}}\big)^{-1}\tau_0\mathcal{F}_{\theta\hat{n}}$ is the new twisted f\/lip operator (\ref{flip}) and the label $\hat{n}$ indicates that the states are $\vec{J}\cdot\hat{n}$ eigenstates with eigenvalue given by the second state label:
\[
\vec{J}\cdot\hat{n}|\nu,\alpha\rangle_{\hat{n}}=\frac{\vec{\sigma}\cdot\hat{n}}{2}|\nu,\alpha\rangle_{\hat{n}}
=\frac{\alpha}{2}|\nu,\alpha\rangle_{\hat{n}},\qquad\alpha=\pm1 .
\]
The f\/irst state label, that is $\nu$ in the above formula, indicates principal quantum number and hence the energy eigenvalue:
\[
P_0|\nu,\alpha\rangle_{\hat{n}}=E_\nu|\nu,\alpha\rangle_{\hat{n}}  .
\]

Likewise we can compute the two-electron excited state
\begin{gather}\label{state}
|2,+1;3,+1\rangle_{\theta\hat{n}}=\frac{\I-\tau_{\theta\hat{n}}}{\sqrt{2}}|2,+1;3,+1\rangle_{\hat{n}}.
\end{gather}
We consider the case where two of the four electrons in Be are initially in the ground state and the remaining two in the state (\ref{state}). Then in the standard $\theta=0$ case, any transition from the excited state~(\ref{state}) to the ground state vanishes by Pauli principle as the ground state is fully occupied. Once the explicit expressions for the two states are obtained it is possible to compute the transition rate in the $\theta\neq0$ case. In the noncommuativity case, the twisted statistics introduces a $\theta\hat{n}$ dependence in energy eigenstates. If we take into account the rotation around its axis and revolution around the sun, the Earth frame is a non-inertial one. This ref\/lects in the fact that in this frame the angular momentum generators rotate. Ef\/fectively, such a rotation can be seen on the axis identif\/ied by $\hat{n}$:
\begin{gather}\label{rot}
\hat{n}^i\to\hat{m}^i:=\hat{n}^jR(t)^i_j,\qquad R(t)\in SO(3) .
\end{gather}
Hence the twisted f\/lip operator goes into $\tau_{\theta\hat{m}}$. Such a change, $\tau_{\theta\hat{n}}\to\tau_{\theta\hat{m}}$ induces (twisted) bosonic components in multi-fermion (in the case of study electrons) state vectors leading to non-Pauli ef\/fects.

The time scale of the ef\/fect induced by the twisted f\/lip operator in energy eigenstates is expected to be very large, of the order of the inverse of the Planck length.  On the other side the ef\/fect of Earth rotation, as in (\ref{rot}), are at the most of the order of 1 year. A suitable approach to treat corrections due by such a short-scale ef\/fect is the {\it sudden approximation}, in the case under study is the rotation $R(t)$ of $\hat{n}$ which is considered as sudden ef\/fect. As $\hat{n}$ rotates, $\tau_{\theta\hat{n}}$~goes into $\tau_{\theta\hat{m}}$ but the states have ``no time'' to adjust, they in fact do not change in the sudden approximation. The states are then no longer eigenstates of the twisted f\/lip opera\-tor~$\tau_{\theta\hat{m}}$, this is the cause of the appearance of non-Pauli ef\/fects. The details of the calculation can be found in~\cite{Pramod}. Comparison between the nonvanishing transition rate in the $\theta\neq0$ case and the bounds on Pauli-forbidden transitions coming from nuclear physics \cite{expe1,expe2,expe3,expe4,expe5} leads to the bound for the energy scale of the noncommutativity parameter $\gtrsim10^{24}$~TeV given above.

\subsection{Particle physics phenomenology}

The noncommutativity parameter $\theta$ can also be constrained using particle physics experiments. Non-commutative spacetimes lead to CPT violation \cite{CPT}. One of the strongest experimental supports to CPT symmetry comes from the mass dif\/ference of $K^0$ and $\overline{K}^0$ which is predicted to vanish exactly for CPT-invariant theories. In~\cite{anosh}, the data from the KTeV E731 experiment~\cite{Fermi} and the experiments on kaons \cite{Kaon1,Kaon2} have been used to get bounds on $\theta$. The constraints obtained are similar to the ones coming from nuclear experiments described above. Below we report the main results.

Considering QFT on the Moyal plane, that is the standard point-wise product is deformed into (\ref{sig1}), we get the following expression for the $K^0$--$\overline{K}^0$ mass dif\/ference at f\/irst order in $\vec{\theta}^0\equiv(\theta^{01},\theta^{02},\theta^{03})$ \cite{anosh}:
\begin{gather}\label{sig28}
m_{K_0}-m_{\bar{K}_0}\simeq\delta_{\perp}\big(m_{K_0}\vec{\theta}^0\vec{P}_{in}\big)\sqrt{1+\tan^2(\phi_{SW})},
\end{gather}
where $\phi_{SW}$ is called the {\it super-weak angle}, while $\delta_{\perp}$ is given in terms of the CPT violating parameter $\delta$ and $\phi_{SW}$: $\delta_\perp=-{\rm Re}\ \delta\ \sin(\phi_{SW})+{\rm Im}\ \delta\ \cos(\phi_{SW})$.

From the KTeV E731 experiment, we get for $\phi_{SW}$:
\begin{gather}\label{sig26}
\phi_{SW}=43.4^\circ\pm0.1^\circ.
\end{gather}
Using the measurements of the complex and real part of $\delta$ from kaon decay experiments \cite{Kaon1, Kaon2} and the above value of $\phi_{SW}$ in~(\ref{sig26}), we get:
\begin{gather}\label{sig27}
\delta_\perp\simeq20.93\ \times 10^{-5} .
\end{gather}

Plugging back (\ref{sig26}) and (\ref{sig27}) in (\ref{sig28}), we can solve for $\vec{\theta}^0$ f\/inding
\[
\sqrt{|\vec{\theta}^0|}<10^{-32}~{\rm m} .
\]
This corresponds to a lower bound for the energy scale associated to noncommutativity of the order of $10^{16}$~GeV.

We emphasis that the above bounds are obtained for ``electric'' noncommutativity ($\vec{\theta}^0\neq0$, $\theta^{ij}\equiv0$) and not for ``magnetic'' noncommutativity ($\vec{\theta}^0\equiv0$, $\theta^{ij}\neq0$ for some~$i$,~$j$).

Further details can be found in~\cite{anosh}.

In \cite{anosh}, CPT measurements on the $g-2$ dif\/ference of $\mu^+$ and $\mu^-$ \cite{Mu1,Mu2,Mu3,Mu4} have been also used to constrain $\theta$. The derivation is similar to the kaon case but, the mass of the muon being considerably smaller than the kaon mass, we get a much weaker bound:
\[
\sqrt{\theta}<10^{-20}~{\rm m} .
\]
It corresponds to a lower bound for the energy scale of $10^3$~GeV.

The measurements of the $g-2$ dif\/ference of~$e^+$ and~$e^-$ are much more precise then the ones on muons. But still they do not give good bounds on~$\theta$, the electron being even lighter than the muon.

\section{Final remarks}\label{sectionVIII}

We can now brief\/ly outline how to generalize our considerations on the Wick--Voros twist (\ref{UV9}) to $2N$-dimensions\footnote{In 2$N+1$-dimensions, we can always choose $\theta_{\mu\nu}$ so that $\theta_{\mu,2N+1}=\theta_{2N+1,\mu}$=0.}. We can always choose $\hat{x}_{\mu}$ so that $\theta_{\mu\nu}$, now an $2N\times 2N$ skew-symmetric matrix, becomes a direct sum of $N$ $2\times 2$ ones.  These $2\times2$ matrices are of the form (\ref{UV10}), but dif\/ferent 2$\times$2 matrices may have dif\/ferent factors $\theta$. For every such $2\times2$ block, we have a pair of $\hat{x}$'s which can be treated as in the 2-dimensional case above. (Of course there is no twist in any block with a vanishing $\theta$.)

\subsection*{Acknowledgements}

It is a pleasure for Balachandran, Marmo and Martone to thank Alberto Ibort and the Universidad Carlos III de Madrid for their wonderful hospitality and support.

The work of Balachandran and Martone was supported in part by DOE under the grant number DE-FG02-85ER40231 and by the Institute of Mathematical Sciences, Chennai. We thank Professor T.R.~Govindarajan for his wonderful hospitality at Chennai. Balachandran was also supported by the Department of Science and Technology, India.

\pdfbookmark[1]{References}{ref}
\LastPageEnding


\begin{thebibliography}{99}

\footnotesize\itemsep=0pt

\bibitem{Doplicher}
Doplicher S., Fredenhagen  K., Roberts J.E.,
Spacetime quantization induced by classical gravity,
\href{http://dx.doi.org/10.1016/0370-2693(94)90940-7}{{\it Phys. Lett.~B}} {\bf 331} (1994), 33--44.

\bibitem{String}
 Seiberg N., Witten E.,
 String theory and noncommutative geometry,
\href{http://dx.doi.org/10.1088/1126-6708/1999/09/032}{{\it J. High Energy Phys.}} {\bf 1999} (1999), no.~9, 032, 93~pages,
\href{http://arxiv.org/abs/hep-th/9908142}{hep-th/9908142}.

\bibitem{DFR2}
Doplicher S., Fredenhagen K., Roberts J.E.,
The quantum structure of spacetime at the Planck scale and quantum f\/ields,
\href{http://dx.doi.org/10.1007/BF02104515}{{\it Comm. Math. Phys.}} {\bf 172} (1995), 187--220, \href{http://arxiv.org/abs/hep-th/0303037}{hep-th/0303037}.



\bibitem{drinfeld}
Drinfel'd V.G.,
Quasi-Hopf algebras,
{\it Leningrad Math. J.} {\bf 1}  (1990), 1419--1457.

\bibitem{chaichian}
 Chaichian M., Kulish P.P., Nishijima K., Tureanu A.,
 On a Lorentz-invariant interpretation of noncommutative space-time and its implications on noncommutative QFT,
\href{http://dx.doi.org/10.1016/j.physletb.2004.10.045}{{\it Phys. Lett. B}} {\bf 604} (2004), 98--102,
\href{http://arxiv.org/abs/hep-th/0408069}{hep-th/0408069}.

\bibitem{wess}
  Wess J.,
  Deformed coordinates spaces: derivatives,
  in Mathematical, Theoretical and Phenomenological Challenges Beyond the Standard Model (Vrnjacka Banja, Serbia, 2003), Editors G.~Djordjevic, L.~Nesic and J.~Wess, World Scientif\/ic, 2005, 122--128,
  \href{http://arxiv.org/abs/hep-th/0408080}{hep-th/0408080}.


\bibitem{sasha}
 Balachandran A.P.,  Pinzul A., Quereshi B.A.,
 Twisted Poincar\'e invariant quantum f\/ield theories,
 \href{http://dx.doi.org/10.1103/PhysRevD.77.025021}{{\it Phys. Rev. D}} {\bf 77} (2008), 025021, 9 pages,
\href{http://arxiv.org/abs/0708.1779}{arXiv:0708.1779}.


\bibitem{Dito}
  Dito G., Sternheimer D.,
  Deformation quantization: genesis, developments and metamorphoses,
   in  Deformation Quantization (Strasbourg, 2001), Editor G.~Halbout,
   {\it IRMA Lect. Math. Theor. Phys.}, Vol.~1, de Gruyter, Berlin, 2002, 9--54, \href{http://arxiv.org/abs/math.QA/02201168}{math.QA/02201168}.

\bibitem{chari}
 Chari V., Pressley A.,
 A guide to quantum groups,
 Cambridge University Press, Cambridge, 1994.

\bibitem{majid}
 Majid S.,
  Foundations of quantum group theory,
    Cambridge University Press, Cambridge, 1995.

\bibitem{aschieri}
 Aschieri P.,
  Lectures on Hopf algebras, quantum groups and twists,
 \href{http://arxiv.org/abs/hep-th/0703013}{hep-th/0703013}.

\bibitem{Lizzi}
 Galluccio S., Lizzi F., Vitale P.,
  Twisted noncommutative f\/ield theory with the Wick--Voros and Moyal products,
  	\href{http://dx.doi.org/10.1103/PhysRevD.78.085007}{{\it Phys. Rev. D}} {\bf 78} (2008), 085007, 14 pages,
  \href{http://arxiv.org/abs/0810.2095}{arXiv:0810.2095}.

\bibitem{Vitale}
 Aschieri P., Lizzi F., Vitale P.,
 Twisting all the way: from classical mechaincs to quantum f\/ields,
 \href{http://dx.doi.org/10.1103/PhysRevD.77.025037}{{\it Phys. Rev.~D}} {\bf 77} (2008), 025037, 16 pages,
  \href{http://arxiv.org/abs/0708.3002}{arXiv:0708.3002}.

\bibitem{bondia}
 Bahns D., Doplicher S., Fredenhagen K., Piacitelli G.,
 On the unitarity problem in space-time noncommutative theories,
 \href{http://dx.doi.org/10.1016/S0370-2693(02)01563-0}{{\it Phys. Lett. B}} {\bf 533} (2002), 178--181,
\href{http://arxiv.org/abs/hep-th/0201222}{hep-th/0201222}.

\bibitem{mangano}
 Balachandran A.P., Govindarajan T.R., Mangano G., Pinzul A., Quereshi B.A., Vaidya S.,
 Statistics and UV-IR mixing with twisted Poincar\'e invariance,
 \href{http://dx.doi.org/10.1103/PhysRevD.75.045009}{{\it Phys. Rev. D}}
 {\bf 75} (2007), 045009, 7 pages,
 \href{http://arxiv.org/abs/hep-th/0608179}{hep-th/0608179}.

\bibitem{Mario}
 Balachandran A.P., Martone M.,
  Twisted quantum f\/ields on Moyal and Wick--Voros planes are inequivalent,
\href{http://dx.doi.org/10.1142/S0217732309031156}{{\it Modern Phys. Lett. A}} {\bf 24} (2009), 1721--1730,
  \href{http://arxiv.org/abs/0902.1247}{arXiv:0902.1247}.

\bibitem{Mario3}
 Balachandran A.P., Ibort A., Marmo G., Martone M.,
  Inequivalence of QFT's on noncommutative spacetimes: Moyal versus Wick--Voros,
  \href{http://dx.doi.org/10.1103/PhysRevD.81.085017}{{\it Phys. Rev. D}} {\bf 81} (2010), 085017, 8 pages
\href{http://arxiv.org/abs/0910.4779}{arXiv:0910.4779}.


\bibitem{Grosse}
Grosse H.,
On the construction of M\"oller operators for the nonlinear Schr\"odigner equation,
\href{http://dx.doi.org/10.1016/0370-2693(79)90835-9}{{\it Phys. Lett. B}} {\bf 86} (1979), 267--271.

\bibitem{Zamo}
  Zamolodchikov A.B., Zamolodchikov Al.B.,
   Factorized $S$-matrices in two dimensions as the exact solutions of certain relativistic quantum f\/ield theory models,
   \href{http://dx.doi.org/10.1016/0003-4916(79)90391-9}{{\it  Ann. Physics}} {\bf 120} (1979), 253--291.

\bibitem{Faddeev}
 Faddeev L.D.,
 Quantum completely integrable models in f\/ield theory,
 in Mathematical Physics Reviews,
  {\it Soviet Sci. Rev. Sect. C: Math. Phys. Rev.}, Vol.~1, Harwood Academic, Chur, 1980, 107--155.

\bibitem{BaPiQur}
 Balachandran A.P., Pinzul A., Qureshi B.A.,
 UV-IR mixing in non-commutative plane,
\href{http://dx.doi.org/10.1016/j.physletb.2006.02.006}{{\it Phys. Lett. B}} {\bf 634} (2006), 434--436,
  \href{http://arxiv.org/abs/hep-th/0508151}{hep-th/0508151}.


\bibitem{BaPiQue}
 Balachandran A.P., Pinzul A., Queiroz A.R.,
  Twisted Poincar\'e invariance, noncommutative gauge theories and UV-IR mixing,
 \href{http://dx.doi.org/10.1016/j.physletb.2008.08.052}{{\it Phys. Lett. B}} {\bf 668} (2008), 241--245,
\href{http://arxiv.org/abs/0804.3588}{arXiv:0804.3588}.


\bibitem{BaIbMaMa}
 Balachandran A.P., Ibort A., Marmo G., Martone M.,
Covariant quantum f\/ields on noncommutative spacetimes, in preparation.


\bibitem{vaidya}
 Balachandran A.P., Mangano G., Pinzul A., Vaidya S.,
 Spin and statistics on the Groenwald--Moyal plane: Pauli-forbidden levels and transitions,
\href{http://dx.doi.org/10.1142/S0217751X06031764}{{\it Internat. J. Modern Phys. A}} {\bf 21} (2006), 3111--3126,
 \href{http://arxiv.org/abs/hep-th/0508002}{hep-th/0508002}.


\bibitem{Mario2}
 Balachandran A.P., Martone M.,
 Spacetime from symmetry: the Moyal plane from the Poincar\'e--Hopf algebra,
\href{http://dx.doi.org/10.1142/S0217732309031144}{{\it  Modern Phys. Lett. A}} {\bf 24} (2009), 1811--1821,
\href{http://arxiv.org/abs/0902.3409}{arXiv:0902.3409}.


\bibitem{gauge1}
 Balachandran A.P., Pinzul A., Qureshi B.A., Vaidya S.,
  Twisted gauge and gravity theories on the Groenewold--Moyal plane,
\href{http://dx.doi.org/10.1103/PhysRevD.76.105025}{{\it Phys. Rev. D}} {\bf 76}  (2007), 105025, 10 pages,
  \href{http://arxiv.org/abs/0902.3409}{arXiv:0708.0069}.

\bibitem{gauge2}
 Balachandran A.P., Pinzul A., Qureshi B.A., Vaidya S.,
  $S$ matrix on the Moyal plane: locality versus Lorentz invariance,
\href{http://dx.doi.org/10.1103/PhysRevD.77.025020}{{\it Phys. Rev. D} {\bf 77}} (2008), 025020, 8 pages,
\href{http://arxiv.org/abs/0902.3409}{arXiv:0708.1379}.

\bibitem{Nonass}
 Balachandran A.P., Qureshi B.A.,
  Poincar\'e quasi-Hopf symmetry and nonassociative spacetime algebra from twisted gauge theories,
\href{http://dx.doi.org/10.1103/PhysRevD.81.065006}{{\it Phys. Rev. D}} {\bf 81} (2010), 065006, 6 pages,
\href{http://arxiv.org/abs/0903.0478}{arXiv:0903.0478}.


\bibitem{NCMB1}
 Chu C.-S., Greene B.R., Shiu G.,
 Remarks on inf\/lation and noncommutative geometry,
\href{http://dx.doi.org/10.1142/S0217732301005680}{{\it  Modern Phys. Lett.~A}} {\bf 16} (2001), 2231--2240,
\href{http://arxiv.org/abs/hep-th/0011241}{hep-th/0011241}.

\bibitem{NCMB2}
 Lizzi F., Mangano G., Miele G., Peloso M.,
  Cosmological perturbations and short distance physics from noncommutative geometry,
\href{http://dx.doi.org/10.1088/1126-6708/2002/06/049}{{\it J. High Energy Phys.}} {\bf 2002}  (2002), no.~6, 049, 16~pages,
  \href{http://arxiv.org/abs/hep-th/0203099}{hep-th/0203099}.

\bibitem{NCMB3}
 Brandenberger R., Ho P.-M.,
  Noncommutative spacetime, stringy spacetime uncertainty principle and density f\/luctuations, \href{http://dx.doi.org/10.1103/PhysRevD.66.023517}{{\it Phys. Rev. D}} {\bf 66}  (2002), 023517, 10~pages,
 \href{http://arxiv.org/abs/hep-th/0203119}{hep-th/0203119}.

\bibitem{NCMB4}
 Huang Q.-G., Li M.,
 CMB power spectrum from noncommutative spacetime,
\href{http://dx.doi.org/10.1088/1126-6708/2003/06/014}{{\it J. High Energy Phys.}} {\bf 2003} (2003), no.~6, 014, 7~pages,
\href{http://arxiv.org/abs/hep-th/0304203}{hep-th/0304203}.


\bibitem{NCMB5}
Huang Q.-G., Li M.,
 Noncommutative inf\/lation and the CMB multipoles,
\href{http://dx.doi.org/10.1088/1475-7516/2003/11/001}{{\it J. Cosmol. Astropart. Phys.}} {\bf 2003} (2003), no.~11, 001, 10~pages,
\href{http://arxiv.org/abs/astro-ph/0308458}{astro-ph/0308458}.

\bibitem{NCMB6}
 Tsujikawa S., Maartens R., Branderberger R.,
 Noncommutative inf\/lation and the CMB,
 \href{http://dx.doi.org/10.1016/j.physletb.2003.09.022}{{\it Phys. Lett. B}} {\bf 574} (2003), 141--148,
\href{http://arxiv.org/abs/astro-ph/0308169}{astro-ph/0308169}.

\bibitem{NCMB7}
 Balachandran A.P., Queiroz A.R., Marques A.M., Teotonio-Sobrinho P.,
 Quantum f\/ields with noncommutative target spaces,
 \href{http://dx.doi.org/10.1103/PhysRevD.77.105032}{{\it Phys. Rev. D}} {\bf 77} (2008), 105032, 11~pages,
\href{http://arxiv.org/abs/0706.0021}{arXiv:0706.0021}.

\bibitem{NCMB8}
 Barosi L., Brito F.A., Queiroz A.R.,
 Noncommutative f\/ield gas driven inf\/lation,
\href{http://dx.doi.org/10.1088/1475-7516/2008/04/005}{{\it J. Cosmol. Astropart. Phys.}} {\bf 2008} (2008), no.~4, 005, 16~pages,
\href{http://arxiv.org/abs/0801.0810}{arXiv:0801.0810}.

\bibitem{NCMB9}
 Fatollahi A.H., Hajirahimi M.,
 Noncommutative black-body radiation: implications on cosmic microwave background,
\href{ http://dx.doi.org/10.1209/epl/i2006-10149-x}{{\it Europhys. Lett.}} {\bf 75} (2006), 542--547,
\href{http://arxiv.org/abs/astro-ph/ 0607257}{astro-ph/ 0607257}.

\bibitem{NCMB10}
 Fatollahi A.H., Hajirahimi M.,
 Black-body radiation of noncommutative gauge f\/ields,
\href{http://dx.doi.org/10.1016/j.physletb.2006.08.081}{{\it Phys. Lett. B}} {\bf 641} (2006), 381--385,
\href{http://arxiv.org/abs/hep-th/0611225}{hep-th/0611225}.

\bibitem{CMB1}
 Akofor E., Balachandran A.P., Jo S.G., Joseph A., Qureshi B.A.,
 Direction-dependent CMB power spectrum and statistical anisotropy from noncommutative geometry, \href{http://dx.doi.org/10.1088/1126-6708/2008/05/092}{{\it J. High Energy Phys.}} {\bf 2008}  (2008), no.~5, 092, 21~pages,
\href{http://arxiv.org/abs/0710.5897}{arXiv:0710.5897}.

\bibitem{CMB2}
 Akofor E., Balachandran A.P., Joseph A., Pekowsky L., Qureshi B.A.,
 Constraints from cosmic microwave background on spacetime noncommutativity and causality violation,
\href{http://dx.doi.org/10.1103/PhysRevD.79.063004}{{\it Phys. Rev. D}} {\bf 79} (2009), 063004, 5~pages,
\href{http://arxiv.org/abs/0806.2458}{arXiv:0806.2458}.


\bibitem{CMBt1}
 Dodelson S.,
 Modern cosmology,
 Academic Press, San Diego, 2003.


\bibitem{CMBt2}
 Mukhanov V.,
  Physical foundations of cosmology,
  Cambridge University Press, New York, 2005.


\bibitem{CMBt3}
 Mukhanov V.F., Feldman H.A., Bradenberger R.H.,
 Theory of cosmological perturbations,
 \href{http://dx.doi.org/10.1016/0370-1573(92)90044-Z}{{\it Phys. Rep.}} {\bf 215} (1992), 203--333.

\bibitem{WMAP1}
 Komatsu E. et al.,
  Five-year Wilkinson microwave anisotropy probe (WMAP) observations: cosmological interpretation,
  \href{http://dx.doi.org/10.1088/0067-0049/180/2/330}{{\it Astrophys. J. Suppl.}} {\bf 180} (2009), 330--376,
\href{http://arxiv.org/abs/0803.0547}{arXiv:0803.0547}.


\bibitem{WMAP2}
 Dunkey J.  et al.,
  Five-year Wilkinson microwave anisotropy probe (WMAP) observations: likelihoods and parameters from the WMAP data,
  \href{http://dx.doi.org/10.1088/0067-0049/180/2/306}{{\it Astrophys. J. Suppl.}} {\bf 180} (2009), 306--329,
 \href{http://arxiv.org/abs/0803.0586}{arXiv:0803.0586}.


\bibitem{WMAP3}
 Nolta M.R.  et al., Five-year Wilkinson microwave anisotropy probe (WMAP) observations: angular power spectra,
  \href{http://dx.doi.org/10.1088/0067-0049/180/2/296}{{\it Astrophys. J. Suppl.}} {\bf 180} (2009), 296--305,
 \href{http://arxiv.org/abs/0803.0593}{arXiv:0803.0593}.


\bibitem{ACb1}
 Reichardte C.L.  et al.,
 High resolution CMB power spectrum from the complete ACBAR data set,
 \href{http://dx.doi.org/10.1088/0004-637X/694/2/1200}{{\it Astrophys.~J.}} {\bf 694} (2009), 1200--1219,
 \href{http://arxiv.org/abs/0801.1491}{arXiv:0801.1491}.


\bibitem{ACb2}
 Kuo C.L.  et al.,
  Improved measurements of the CMB power spectrum with ACBAR,
\href{http://dx.doi.org/10.1086/518401}{{\it Astrophys. J.}} {\bf 664} (2007), 687--701,
 \href{http://arxiv.org/abs/astro-ph/0611198}{astro-ph/0611198}.


\bibitem{ACb3}
 Kuo C.L.  et al.,
  High resolution observations of the CMB power spectrum with ACBAR,
\href{http://dx.doi.org/10.1086/379783}{{\it Astrophys. J.}} {\bf 600} (2004), 32--51,
\href{http://arxiv.org/abs/astro-ph/0212289}{astro-ph/0212289}.


\bibitem{CB1}
 Mason B.S.  et al.,
 The anisotrpy of the microwave backgound to $I=3500$: deep f\/ield observations with the cosmic background imager,
\href{http://dx.doi.org/10.1086/375507}{{\it Astrophys. J.}} {\bf 591} (2007), 540--555,
 \href{http://arxiv.org/abs/astro-ph/0205384}{astro-ph/0205384}.


\bibitem{CB2}
 Sivers J.L.  et al.,
 Implications of the cosmic background imager polarization data,
  {\it Astrophys. J.} {\bf 660} (2007), 976--987,
\href{http://arxiv.org/abs/astro-ph/0509203}{astro-ph/0509203}.


\bibitem{CB3}
 Sivers J.L.  et al.,
  Cosmological parameters from cosmic background imager observations and comparisons with BOOMERANG, DASI, and MAXIMA,
  \href{http://dx.doi.org/10.1086/375510}{{\it Astrophys. J.}} {\bf 591} (2003), 599--622,
\href{http://arxiv.org/abs/astro-ph/0205387}{astro-ph/0205387}.


\bibitem{CB4}
 Pearson T.J.  et al.,
 The anisotropy of the microwave background to $I=3500$: mosaic observations with the cosmic background imager,
 \href{http://dx.doi.org/10.1086/375508}{{\it Astrophys. J.}} {\bf 591} (2003), 556--574,
\href{http://arxiv.org/abs/astro-ph/0205388}{astro-ph/0205388}.


\bibitem{CB5}
 Readhead A.C.S.  et al.,
 Extended mosaic observations with the cosmic background imager,
 \href{http://dx.doi.org/10.1086/421105}{{\it Astrophys. J.}} {\bf 609} (2004), 498--512,
\href{http://arxiv.org/abs/astro-ph/0402359}{astro-ph/0402359}.

\bibitem{pot}
 Chakraborty B., Gangopadhyay S., Hazra A.G., Sholtz F.G.,
  Twisted Galilean symmetry and the Pauli principle at low energies,
\href{http://dx.doi.org/10.1088/0305-4470/39/30/011}{{\it J. Phys. A: Math. Gen.}} {\bf 39} (2006), 9557--9572,
 \href{http://arxiv.org/abs/hep-th/0601121}{hep-th/0601121}.


\bibitem{Pramod}
 Balachandran A.P., Joseph A., Padmanabhan P.,
 Non-Pauli transitions from spacetime noncommutativity,
\href{http://arxiv.org/abs/1003.2250}{arXiv:1003.2250}.\\
  Balachandran A.P., Padmanabhan P.,
 Non-Pauli ef\/fects from noncommutative spacetimes,
 \href{http://arxiv.org/abs/1006.11851}{arXiv:1006.1185}.

\bibitem{expe1}
 Back H.O.  et al. [Borexino collaboration],
  New experimental limits on violations of the Pauli exclusion prin\-cip\-le obtained with the Borexino counting test facility,
\href{http://dx.doi.org/10.1140/epjc/s2004-01991-1}{{\it Eur. Phys. J. C}} {\bf 37} (2004), 421--431,
\mbox{\href{http://arxiv.org/abs/hep-ph/0406252}{hep-ph/0406252}}.

\bibitem{expe2}
 Barabash A.S., Kornoukhov V.N., Tsipenyuk Yu.M., Chapyzhnikov B.A.,
  Search for anomalous carbon atoms -- evidence of violation of the Pauli principle during the period of nucleosynthesis,
  \href{http://dx.doi.org/10.1134/1.567831}{{\it JETP Lett.}} {\bf 68} (1998), 112--116.\\
  Arnold  R.  et al.,
    Testing the Pauli exclusion principle with the NEMO-2 detector,
  \href{http://dx.doi.org/10.1007/s100500050354}{{\it Eur. Phys. J. A}} {\bf 6} (1999), 361--366.


\bibitem{expe3}
 Ramberg E., Snow G.A.,
 Experimental limit on a small violation of the Pauli principle,
 \href{http://dx.doi.org/10.1016/0370-2693(90)91762-Z }{{\it Phys. Lett. B}} {\bf 238} (1990), 438--411.

\bibitem{expe4}
 Suzuki Y.  et al. [Kamiokande collaboration],
 Study of invisible nucleon decay, $N \to \nu\nu\bar\nu$, and a forbidden nuclear transition in the Kamiokande detector,
  \href{http://dx.doi.org/10.1016/0370-2693(93)90582-3}{{\it Phys. Lett. B}} {\bf 311} (1993), 357--361.

\bibitem{expe5}
 Bartalucci S.  et al. [VIP collaboration],
  New experimental limit on the Pauli exclusion principle violation by electrons,
\href{http://dx.doi.org/10.1016/j.physletb.2006.07.054}{{\it Phys. Lett. B}} {\bf 641} (2006), 18--22.

\bibitem{CPT}
Akofor E., Balachandran A.P., Jo S.G., Joseph A.,
Quantum f\/ields on the Groenwold--Moyal plane: $C$, $P$, $T$ and $CPT$,
\href{http://dx.doi.org/10.1088/1126-6708/2007/08/045}{{\it J. High Energy Phys.}} {\bf 2007}  (2007), no.~8, 045, 14~pages,
\href{http://arxiv.org/abs/0706.1259}{arXiv:0706.1259}.

\bibitem{anosh}
Joseph A.,
Particle phenomenology on noncommutative spacetime,
\href{http://dx.doi.org/10.1103/PhysRevD.79.096004}{{\it Phys. Rev. D}} {\bf 79}  (2009), 096004, 9~pages,
\href{http://arxiv.org/abs/0811.3972}{arXiv:0811.3972}.

\bibitem{Fermi}
Gibbons L.K.  et al.,
$CP$ and $CPT$ symmetry tests from the two-pion decays of the neutral kaon with the Fermilab E731 detector,
\href{http://dx.doi.org/10.1103/PhysRevD.55.6625}{{\it Phys. Rev. D}} {\bf 55} (1997), 6625--6715


\bibitem{Kaon1}
Angelopoulos  A. et al. [CPLEAR collaboration],
A determination of the CPT violation parameter ${\rm Re}(\delta)$ from the semileptonic decay of strangeness tagged neutral kaons,
\href{http://dx.doi.org/10.1016/S0370-2693(98)01357-4}{{\it Phys. Lett. B}} {\bf 444} (1998), 52--60.

\bibitem{Kaon2}
Lai  A. et al. [NA48 collaboration],
Search for CP violation in $K^0\to3\pi^0$ decays,
\href{http://dx.doi.org/10.1016/j.physletb.2005.01.065}{{\it Phys. Lett. B}} {\bf 610} (2005), 165--176,
\href{http://arxiv.org/abs/hep-ex/0408053}{hep-ex/0408053}.

\bibitem{Mu1} Bennett  G.W. et al. [Muon (g-2) collaboration],
 Search for Lorentz and $CPT$ violation ef\/fects in muon spin precession,
 \href{http://dx.doi.org/10.1103/PhysRevLett.100.091602}{{\it Phys. Rev. Lett.}} {\bf 100} (2008), 091602, 5~pages,
\href{http://arxiv.org/abs/0709.4670}{arXiv:0709.4670}.

\bibitem{Mu2} Bailey  J.  et al. [CERN--Mainz--Daresbury collaboration],
Final report on the CERN muon storage ring including the anomalous magnetic moment and the electric dipole moment of the muon, and a direct test of relativistic time dilation,
\href{http://dx.doi.org/10.1016/0550-3213(79)90292-X}{{\it Nuclear Phys. B}} {\bf 150} (1979), 1--75.

\bibitem{Mu3} Carey  R.M.  et al.,
New measurement of the anomalous magnetic moment of the positive muon,
\href{http://dx.doi.org/10.1103/PhysRevLett.82.1632}{{\it Phys. Rev. Lett.}} {\bf 82} (1999), 1632--1635.

\bibitem{Mu4} Brown  H.N. et al. [Muon (g-2) collaboration],
Precise measurement of the positive muon anomalous magnetic moment,
\href{http://dx.doi.org/10.1103/PhysRevLett.86.2227}{{\it Phys. Rev. Lett.}} {\bf 86} (2001), 2227--2231,
\href{http://arxiv.org/abs/hep-ex/0102017}{hep-ex/0102017}.

\end{thebibliography}
\end{document}